\documentclass[preprint]{aastex631}

\usepackage{comment}

\begin{document}

\title{The Perturbed Full Two-Body Problem: Application to Post-DART Didymos}

\author[0000-0001-8437-1076]{Alex J. Meyer}
\affiliation{Smead Department of Aerospace Engineering Sciences, University of Colorado Boulder, CO, USA}

\author[0000-0002-3544-298X]{Harrison F. Agrusa}
\affiliation{Universit\'{e} C\^{o}te d'Azur, Observatoire de la C\^{o}te d'Azur, CNRS, Laboratoire Lagrange, Nice, France}
\affiliation{University of Maryland, College Park, MD, USA}

\author{Derek C. Richardson}
\affiliation{University of Maryland, College Park, MD, USA}

\author{R. Terik Daly}
\affiliation{Johns Hopkins University Applied Physics Laboratory, Laurel, MD, USA}

\author[0000-0001-5875-1083]{Oscar Fuentes-Mu\~{n}oz}
\affiliation{Smead Department of Aerospace Engineering Sciences, University of Colorado Boulder, CO, USA}

\author{Masatoshi Hirabayashi}
\affiliation{Auburn University, Auburn, AL, USA}

\author[0000-0002-0884-1993]{Patrick Michel}
\affiliation{Universit\'{e} C\^{o}te d'Azur, Observatoire de la C\^{o}te d'Azur, CNRS, Laboratoire Lagrange, Nice, France}
\affiliation{University of Tokyo, Department of Systems Innovation, School of Engineering, Tokyo, Japan}

\author[0000-0002-5566-0618]{Colby C. Merrill}
\affiliation{Cornell University, Ithaca, NY, USA}

\author{Ryota Nakano}
\affiliation{Auburn University, Auburn, AL, USA}

\author{Andrew F. Cheng}
\affiliation{Johns Hopkins University Applied Physics Laboratory, Laurel, MD, USA}

\author{Brent Barbee}
\affiliation{NASA Goddard Space Flight Center, Greenbelt, MD, USA}

\author{Olivier S. Barnouin}
\affiliation{Johns Hopkins University Applied Physics Laboratory, Laurel, MD, USA}

\author{Steven R. Chesley}
\affiliation{Jet Propulsion Laboratory, California Institute of Technology, Pasadena, CA, USA}

\author{Carolyn M. Ernst}
\affiliation{Johns Hopkins University Applied Physics Laboratory, Laurel, MD, USA}

\author{Ioannis Gkolias}
\affiliation{Aristotle University of Thessaloniki, Greece}

\author{Nicholas A. Moskovitz}
\affiliation{Lowell Observatory, Flagstaff, AZ, USA}

\author{Shantanu P. Naidu}
\affiliation{Jet Propulsion Laboratory, California Institute of Technology, Pasadena, CA, USA}

\author{Petr Pravec}
\affiliation{Astronomical Institute of the Czech Academy of Sciences, Ond\v{r}ejov, Czech Republic}

\author{Petr Scheirich}
\affiliation{Astronomical Institute of the Czech Academy of Sciences, Ond\v{r}ejov, Czech Republic}

\author{Cristina A. Thomas}
\affiliation{Northern Arizona University, Flagstaff, AZ, USA}

\author{Kleomenis Tsiganis}
\affiliation{Aristotle University of Thessaloniki, Greece}

\author{Daniel J. Scheeres}
\affiliation{Smead Department of Aerospace Engineering Sciences, University of Colorado Boulder, CO, USA}



\begin{abstract}
With the successful impact of the NASA DART spacecraft in the Didymos-Dimorphos binary asteroid system, we provide an initial analysis of the post-impact perturbed binary asteroid dynamics. To compare our simulation results with observations, we introduce a set of ``observable elements'' calculated using only the physical separation of the binary asteroid, rather than traditional Keplerian elements. Using numerical methods that treat the fully spin-orbit-coupled dynamics, we estimate the system's mass and the impact-induced changes in orbital velocity, semimajor axis, and eccentricity. We find that the changes to the mutual orbit depend strongly on the separation distance between Didymos and Dimorphos at the time of impact. If Dimorphos enters a tumbling state after the impact, this may be observable through changes in the system's eccentricity and orbit period. We also find that any DART-induced reshaping of Dimorphos would generally reduce the required change in orbital velocity to achieve the measured post-impact orbit period and will be assessed by the ESA Hera mission in 2027.

\end{abstract}

\keywords{}


\section{Introduction} \label{sec:intro}
In the first planetary defense test of a kinetic impactor, the NASA Double Asteroid Redirection Test (DART) spacecraft impacted Dimorphos, the secondary in the Didymos binary asteroid system, on September 26, 2022 \citep{daly2023}. The impact altered the trajectory of Dimorphos around Didymos, reducing the orbit period by around 33 minutes \citep{thomas2023}. Initial analysis of observations of the system reveals a reduction in the secondary's tangential (along-track) component of orbital velocity by about 2.7 mm/s \citep{cheng2023}. This corresponds to a momentum enhancement factor in the range of 2.2--4.9, depending on the unmeasured mass of Dimorphos, indicating the ejecta launched by the impact had a larger contribution to the change in momentum than the actual DART impact itself \citep{cheng2023}. The analysis by \cite{cheng2023} serves as a good first look into the post-impact dynamics of Didymos, which will be measured accurately in detail by the ESA Hera mission in 2027 \citep{michel2022esa}. However, \cite{cheng2023} only provides a high-level analysis of the impact-induced change in velocity and does not document fully the post-impact orbit dynamics. In this work, we expand on their analysis to characterize in detail the post-impact orbit by calculating the resulting changes to the system's mutual semimajor axis and eccentricity. We also explore how perturbations affect a binary asteroid orbit in general.

Near-Earth binary asteroids usually experience significant spin-orbit coupling due to the proximity and irregular shapes of the two bodies, a configuration commonly referred to as the full two-body problem (F2BP). Equilibrium states of the F2BP have been studied extensively in the literature \citep{scheeres2006relative,bellerose2008energy,jacobson2011long,mcmahon2013dynamic,moeckel2018counting}, but there are few studies on the perturbed dynamics. Of the studies of perturbed binary dynamics, many focus on the rotational dynamics of the secondary: \cite{mcmahon2013dynamic}, \cite{wang2021break}, \cite{naidu2015near}, and \cite{jafari2023surfing} in two dimensions and \cite{agrusa2021excited}, \cite{cuk2021barrel}, \cite{quillen2022non}, and \cite{tan2023attitude} in three dimensions. However, studies on the mutual orbit dynamics of perturbed systems are lacking. \cite{fahnestock2008simulation} simulated the near-Earth binary asteroid (66391) Moshup (formerly 1999 KW4) in a non-equilibrated, non-planar orbit. \cite{cuk2010orbital} also investigated how secondary libration affects the mutual orbit of a binary system. Besides these, the majority of work on perturbed binary asteroid orbit dynamics was done in preparation for the DART impact \citep{meyer2021libration,richardson2022predictions}. What the literature lacks is a comprehensive outline of the mutual orbit dynamics of a system perturbed out of an equilibrium, which we attempt to provide here.

Following previous analyses of the Didymos system and DART impact, we model the fully coupled dynamics using the General Use Binary Asteroid Simulator (\textsc{gubas}) \citep{davis2020doubly}, which has been benchmarked against similar F2BP solvers \citep{agrusa2020benchmarking,ho2023accuracy}. \textsc{gubas} uses inertia integrals in a recursive algorithm to efficiently calculate the mutual potential and its derivatives between two arbitrary rigid bodies \citep{hou2017mutual}. Recently, \textsc{gubas} was used to estimate the DART impact's momentum enhancement factor \citep{cheng2023}, and we will follow a similar approach as that work. We will provide a thorough dynamical analysis of the expected changes to the Didymos system's mutual orbit as a result of the DART impact.

The main contributions of this work are as follows. We introduce a set of so-called `observable elements' to help bridge the gap between observations of binary asteroids and numerical models. We also provide a general discussion of perturbed binary asteroid dynamics in which we define analytical expressions for a perturbation sufficient enough to push a binary asteroid out of its equilibrium configuration, and discuss the applicability of the classical averaged Lagrangian Planetary Equations to both an equilibrated and perturbed binary asteroid. Specifically to Didymos and the DART impact, we expand upon the work by \cite{daly2023} and \cite{cheng2023} in their calculation of the system's mass and the impact-induced change in orbital velocity, respectively. We then calculate the resulting change in mutual semimajor axis and eccentricity, which have not yet been discussed in the literature. We also discuss how secondary attitude instabilities, mass loss, and reshaping can affect the mutual orbit dynamics of the post-impact Didymos system.

This paper first presents a general discussion of the F2BP using both numerical and analytical techniques in Section \ref{sec:f2bp}. In Section \ref{sec:dart}, we present our application to the DART impact and show how the mutual orbit was changed as a result of the impact. In Section \ref{sec:tumbling}, we discuss the implications of a tumbling secondary on the orbit dynamics. We then discuss the implications of Dimorphos mass loss and reshaping in Section \ref{sec:reshaping}. We conclude in Section \ref{sec:conclusion}.

\section{The Full Two-Body Problem} \label{sec:f2bp}

\subsection{Problem Setup}
The latest system parameters for the pre- and post-impact Didymos system, which we use in our numerical simulations, are given in Table \ref{parameters}. In calculating the axis ratios for Didymos and Dimorphos, we use the shape extents defined in \cite{daly2023}. For a conservative approach, we triple the uncertainties in the extents to ensure we sample the full $3\sigma$ parameter space. We limit a/b $>1.01$ for both shapes to avoid cases where b$>$a, which corresponds to an unstable pre-impact equilibrium \citep{bellerose2008energy}, or a$=$b, in which the system decouples and the spin-orbit equilibrium vanishes. Following \cite{cheng2023}, we employ a suite of Monte Carlo simulations that sample over the uncertainties in the pre- and post-impact orbits as well as the uncertainties in the body shapes.

Note the considerable uncertainty on the pre-impact semimajor axis, which is equal to the separation between Didymos and Dimorphos at the impact time under the circular orbit assumption. While DART imaged the system before the impact, there are significant uncertainties in the positions of the bodies' centers of mass since the internal density distributions are unknown, which manifests in a large uncertainty in the pre-impact semimajor axis. Thus, the pre-impact separation is instead measured with radar data in \cite{thomas2023}. This is equal to the pre-impact semimajor axis in the circular pre-impact orbit assumption.

\begin{table}[ht]
\caption{\label{parameters} Parameters for the Didymos binary asteroid system. Uncertainties are reported as $1\sigma$ Gaussian unless noted as a uniform distribution. The secondary diameter and axis ratios are from the pre-impact body, and truncated at a lower limit of a/b $=1.01$.}
\begin{center}
\begin{tabular}{ |c|c|c|c| } 
\hline
Parameter & Value & Uncertainty/Note & Source \\
 \hline
Pre-impact orbit period [hr] & 11.92148 & $\pm0.000044$ & \cite{naidu2022anticipating}  \\
Post-impact orbit period [hr] & 11.372 & $\pm0.0057$ & \cite{thomas2023} \\
Pre-impact semimajor axis [m] & 1206  & $\pm35$ & \cite{thomas2023}  \\
Pre-impact eccentricity & 0 & \textit{assumed} & \cite{naidu2022anticipating}  \\
Primary diameter [m] & 761 & $\pm26$ & \cite{daly2023}  \\
Secondary diameter [m] & 151 & $\pm5$ & \cite{daly2023}  \\
Primary a/b axis ratio & 1.01-1.11 & \textit{uniform} ($3\sigma$)  & \cite{daly2023}  \\
Primary b/c axis ratio & 1.21-1.56  & \textit{uniform} ($3\sigma$) & \cite{daly2023}  \\
Secondary a/b axis ratio & 1.01-1.13  & \textit{uniform} ($3\sigma$) & \cite{daly2023}  \\
Secondary b/c axis ratio & 1.32-1.69  & \textit{uniform} ($3\sigma$)  & \cite{daly2023}  \\
Secondary density [kg/m$^3$] & 1500-3300 & \textit{uniform} ($3\sigma$) & \cite{daly2023}  \\
 \hline
\end{tabular}
\end{center}
\end{table}

In the simulations, we first numerically determine the required density of the primary to achieve an equilibrated randomly selected pre-impact system in a manner similar to previous work \citep{agrusa2021excited, meyer2021libration, cheng2023}. We then iterate an instantaneous $\Delta \vec{v}$ on the secondary's orbital velocity until the system converges to the post-impact orbit period. We include radial and out-of-plane $\Delta \vec{v}$ components such that the full $\Delta \vec{v}$ vector is anywhere within $30^\circ$ of the orbit tangent direction, consistent with the impact and ejecta geometries discussed in \cite{daly2023} and \cite{cheng2023}, respectively. One difference between our algorithm and that of \cite{cheng2023} is how we sample the asteroid shapes. Rather than only using the shape extents provided by \cite{daly2023}, we first sample the volume-equivalent diameter, then determine the ellipsoidal shape by sampling the axis ratios. While the axis ratios are calculated from the shape extents in \cite{daly2023}, the volume of Dimorphos is always self-consistent with the estimated volumes.

We adopt many of the same assumptions as \cite{cheng2023}. Namely, we assume uniform density distributions of the asteroids and an equilibrated pre-impact orbit with zero eccentricity and zero secondary libration. \cite{richardson2022predictions} and \cite{cheng2023} outline justifications for these assumptions. In this work we ignore any torque applied to Dimorphos from the impact, as we are mainly concerned with the orbital dynamics. However, future studies considering the libration and stability of the secondary should include this quantity. We also examine the effects of potential mass loss and reshaping of Dimorphos, whereas \cite{cheng2023} assumed no mass loss or reshaping.

In our \textsc{gubas} simulations, both Didymos and Dimorphos are modeled as triaxial ellipsoids with semiaxes a $\ge$ b $\ge$ c. We use a second-degree and -order gravity expansion between these bodies. Given the uncertainties in the body shapes and their unknown internal mass distributions, there is no advantage to using a higher-order gravity expansion. 

\subsection{Equilibrium Dynamics}
The full two-body problem (F2BP) is driven by spin-orbit coupling, where the bodies' spins and the orbit dynamics are inseparable \citep{duboshin1958differential,maciejewski1995reduction}. The equilibrium configurations of the F2BP have been studied extensively \citep{scheeres2006relative,scheeres2009stability,mcmahon2013dynamic,moeckel2018counting}. In the minimum-energy stable equilibrium configuration, the minor principal axes of both bodies are aligned and the orbit rate is constant. For our purposes we are concerned with the singly-synchronous configuration, where the primary is rotating rapidly and the secondary is tidally locked with a constant orbit rate, which is common among near-Earth binary asteroids \citep{pravec2016binary}. While this is not the true, minimum-energy doubly-synchronous equilibrium, the rapid rotation of the primary tends to decouple its spin from the system and the singly-synchronous state can be considered an equilibrium state. This is equivalent to reducing the problem to the case where the primary is axisymmetric. The equilibrium orbit rate of such a system is \citep{scheeres2009stability,mcmahon2013dynamic}
\begin{equation}
    \dot{\theta}^*=\sqrt{\frac{\mu}{r^3}\left(1+\frac{3(\bar{I}_{A,z}-\bar{I}_{A,xy}-2\bar{I}_{B,x}+\bar{I}_{B,y}+\bar{I}_{B,z})}{2r^2}\right)}
    \label{f2bp_eq}
\end{equation}
for separation distance $r$, where the bar indicates the mass-normalized principal moments of inertia, the subscript $A$ refers to the primary and $B$ to the secondary, and $x$, $y$, and $z$ refer to the minimum, intermediate, and major principal axes. For the axisymmetric primary in our analytic model, $xy$ is used as there is no difference between $x$ and $y$. While the equilibrium is a physically circular orbit (i.e., the separation distance $r$ does not change), the orbit rate is not equal to that of a circular Keplerian orbit. Thus, in the equilibrium F2BP the secondary is trapped at either periapsis or apoapsis while the Keplerian elliptical orbit precesses at the equilibrium orbit rate \citep{scheeres2009stability}. Specifically, if $\bar{I}_{A,z}+\bar{I}_{B,y}+\bar{I}_{B,z}>\bar{I}_{A,xy}+2\bar{I}_{B,x}$, then $\dot{\theta}^*>\sqrt{\mu/r^3}$, the secondary is trapped at periapsis, and the orbit precesses faster than the Keplerian circular orbit rate. Alternatively, if $\bar{I}_{A,z}+\bar{I}_{B,y}+\bar{I}_{B,z}<\bar{I}_{A,xy}+2\bar{I}_{B,x}$, we have the opposite case: the secondary is trapped at apoapsis and the orbit precesses slower than the Keplerian circular orbit rate. In this work, we focus on the former, where the secondary is trapped at periapsis, but in either case, the true anomaly is constant (equal to 0 or $\pi$) while the argument of periapsis tracks the secondary. This implies the orbit will have non-zero Keplerian eccentricity at equilibrium, despite being physically circular. For a system trapped at periapsis, the equilibrium Keplerian eccentricity is calculated from the orbit angular momentum and energy as \citep{scheeres2009stability,mcmahon2013dynamic}
\begin{equation}
    e^*_{\text{Kep}} = \frac{3(\bar{I}_{A,z}-\bar{I}_{A,xy}-2\bar{I}_{B,x}+\bar{I}_{B,y}+\bar{I}_{B,z})}{2r^2},
    \label{kep_ecc}
\end{equation}
while the equilibrium Keplerian semimajor axis is \citep{scheeres2009stability,mcmahon2013dynamic}
\begin{equation}
    a^*_{\text{Kep}} = r\left(1-\frac{3(\bar{I}_{A,z}-\bar{I}_{A,xy}-2\bar{I}_{B,x}+\bar{I}_{B,y}+\bar{I}_{B,z})}{2r^2}\right)^{-1}.
\end{equation}
This semimajor axis is larger than the separation distance in the physically circular orbit owing to the spin-orbit coupling as long as $\bar{I}_{A,z}+\bar{I}_{B,y}+\bar{I}_{B,z}>\bar{I}_{A,xy}+2\bar{I}_{B,x}$, i.e. the secondary is trapped at periapsis.

\subsection{Observable Elements}
Since the Keplerian semimajor axis and eccentricity are not accurate descriptions of what an external observer would see, it is convenient to define so-called ``observable elements'' which are calculated using only the separation distance of the mutual orbit over an averaging window \citep{meyer2021modeling}. These elements serve a similar purpose as geometric orbit elements \citep{borderies1994test,renner2006use} but are distinct as they are intended to be analogous to real-world observations. Furthermore, geometric elements make no consideration for the non-spherical shape of the secondary, which is important in binary asteroid dynamics. The observable semimajor axis is defined as
\begin{equation}
    a_{\mathrm{obs}} = \frac{R_{a,t}+R_{p,t}}{2}
\end{equation}
where $R_{a,t}$ is the maximum separation of the two asteroids over a time period $t$ and similarly $R_{p,t}$ is the minimum separation over the same time. The observable eccentricity is defined in a similar fashion:
\begin{equation}
    e_{\mathrm{obs}} = \frac{R_{a,t}-R_{p,t}}{R_{a,t}+R_{p,t}}.
\end{equation}
These observable elements are more indicative of the physical behavior of the system, with the drawback of being more similar to an averaged element than an osculating Keplerian element, as multiple data points are required to calculate one instance of the observed element. To calculate the observable elements, we use a window approximately equal to the average orbit period. Henceforth, we will refer to the observable semimajor axis and eccentricity as $a$ and $e$ without subscripts, respectively. Keplerian elements will be referred to with the subscript ``Kep".

\begin{figure}[ht!]
   \centering
   \includegraphics[width = 3in]{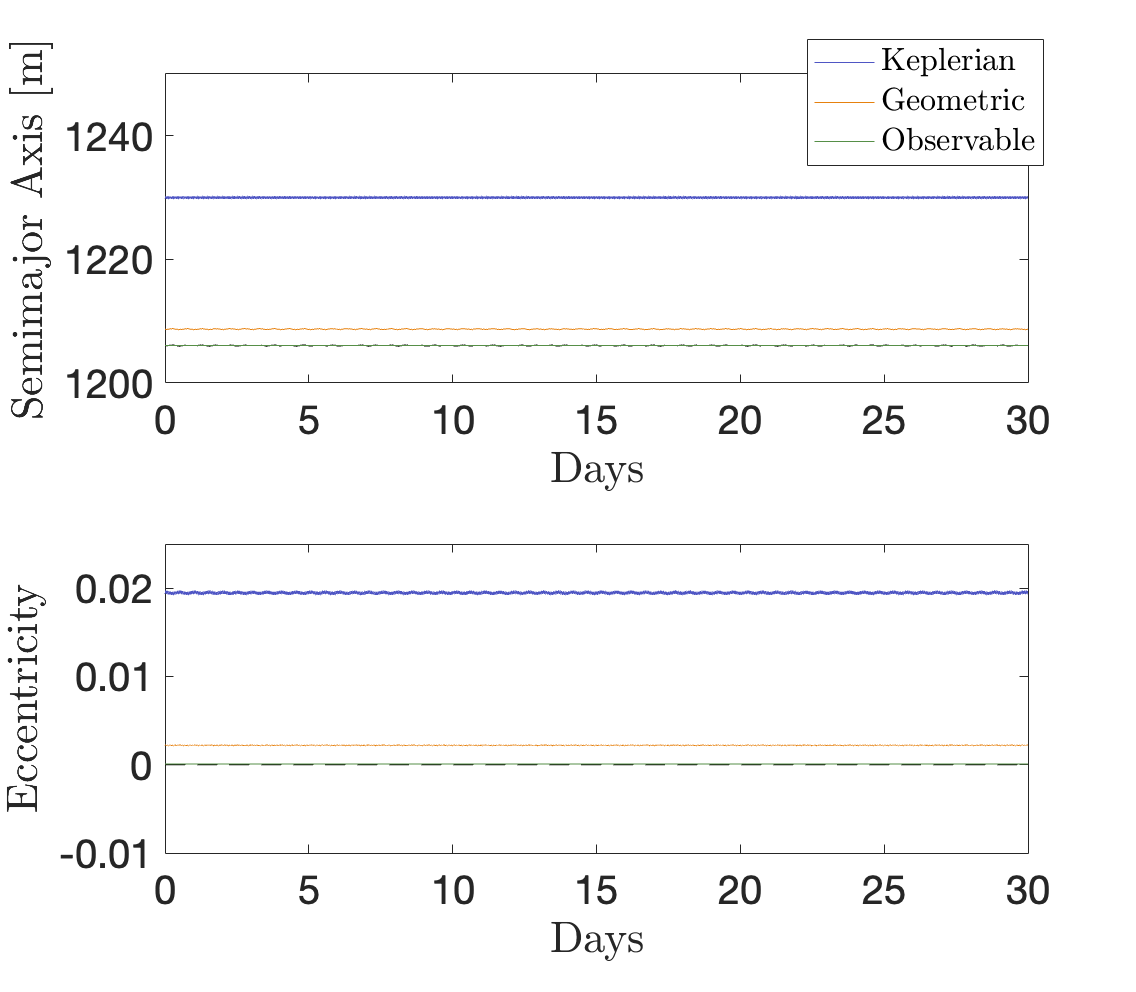} 
   \caption{A comparison between Keplerian, geometric, and observable elements for an equilibrated binary asteroid, showing the semimajor axis (top) and eccentricity (bottom). The physical parameters are shown as a black dashed line. While geometric elements are an improvement over Keplerian, the observable elements do the best job of describing the physical system.}
   \label{element_compare}
\end{figure}

To illustrate the utility of these observable elements, we compare Keplerian, geometric, and observable elements for an equilibrated binary asteroid. Fig. \ref{element_compare} shows this comparison for the semimajor axis and eccentricity. Keep in mind, the equilibrium configuration has a constant separation distance between the primary and secondary. Fig. \ref{element_compare} shows how the Keplerian elements give an eccentric orbit with a semimajor axis larger than the separation distance. The geometric elements are an improvement, as they take into account the oblateness of the primary, but their lack of consideration of the secondary's shape also leads to some errors. The observable elements show an observable semimajor axis equal to the separation distance as the observable eccentricity is equal to 0.

The departure from Keplerian dynamics means Kepler's third law is no longer applicable and the definition of the orbit period becomes ambiguous for close binary asteroids. Following the theme of paralleling observations, we define the ``stroboscopic orbit period'' as the time between successive crossings of the secondary through an arbitrary inertial plane with normal vector perpendicular to the system's total angular momentum \citep{meyer2021libration}. This mimics real-world lightcurve observations, which calculate the orbit period using mutual event timings. In an equilibrium configuration, this period is constant and can be calculated analytically using Eq.\ \ref{f2bp_eq}. However, when the system is perturbed, the stroboscopic orbit period can fluctuate, driven by the precession of the orbit and the libration of the secondary \citep{meyer2021libration}.

We note that the estimated pre-impact semimajor axis and eccentricity values reported in Table \ref{parameters} correspond to the observable semimajor axis and observable eccentricity. The purpose of these observable elements is to facilitate the conversion between observations--which can only deal with the physical positions of the asteroids--and numerical simulations.

\subsection{Perturbed Dynamics}
While the F2BP equilibrium has been studied extensively in past work, we focus on perturbations to this equilibrium. Specifically, there is a transition point where the true anomaly switches from librating about zero to actually tracking the secondary. At this point there is a sharp decrease in the rate of precession of the argument of periapsis where it abruptly changes from precessing at the orbit rate to precessing at a rate dominated by the oblateness of the primary and elongation of the secondary \citep{fahnestock2008simulation,borderies1990phobos}. Here we calculate the perturbation to the orbit necessary to reach this transition point.

The equilibrium spin rate corresponds to an equilibrium specific orbital angular momentum:
\begin{equation}
    h^*=r^2\dot{\theta}^*.
    \label{angmntm}
\end{equation}

\begin{figure}[ht!]
   \centering
   \includegraphics[width = 3in]{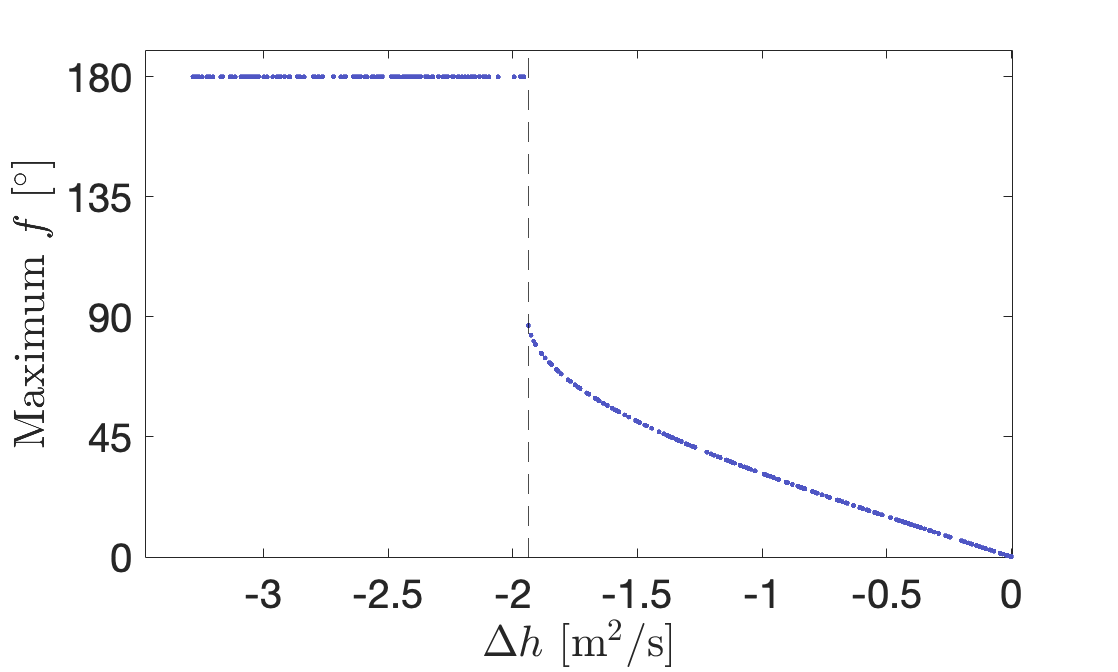} 
   \caption{The maximum true anomaly as a function of the change in specific orbital angular momentum for an orbit perturbation. There is a discontinuity where the perturbation to the orbit changes the orbital angular momentum enough to push the system out of an equilibrium and the true anomaly switches from librating to circulating.}
   \label{deltah_decrease}
\end{figure}

To calculate the perturbation needed to push the system out of equilibrium and allow the true anomaly to track the secondary, we can substitute the mean motion into a perturbed form of Eq \ref{angmntm}:
\begin{equation}
    h^*+\Delta h=r^2\sqrt{\frac{\mu}{r^3}}
    \label{equilibrium}
\end{equation}
and solve for the perturbation $\Delta h$:
\begin{equation}
    \Delta h=r^2\sqrt{\frac{\mu}{r^3}}-h^*.
    \label{analytic_perturb1}
\end{equation}
We compare this analytic result to numerical \textsc{gubas} simulations of an orbit perturbation to an equilibrated Didymos. The numerical simulations begin with the nominal Didymos system, ignoring any uncertainties in Table \ref{parameters} for now, then apply a tangential $\Delta v_T$ to perturb the orbit. Over the simulation time we record the maximum true anomaly value. The results are shown in Fig.\ \ref{deltah_decrease}, where the maximum true anomaly $f$ is plotted as a function of the perturbation $\Delta h$. As the magnitude of $\Delta h$ increases, the maximum allowable true anomaly increases. Outside of a perfect equilibrium where $f=0$, the true anomaly oscillates around 0 until the critical perturbation pushes the maximum true anomaly over $90^\circ$, allowing it to circulate and track the secondary. At this point there is a discontinuity where the maximum true anomaly is $180^\circ$. In Fig.\ \ref{deltah_decrease}, the vertical dashed line corresponds to the analytic value calculated from Eq.\ \ref{analytic_perturb1} using the nominal Didymos system (ignoring uncertainties). We see excellent agreement between this analytic value and the numerical simulations, which are plotted as individual data points.

This perturbation corresponds to decreasing the orbit period, but it is also possible to perturb the orbit out of an equilibrium by increasing the orbit period. This approach requires more consideration. Note that setting $\dot{\theta}=\sqrt{\frac{\mu}{r^3}}$ is equivalent to decreasing the semimajor axis from the equilibrium value $a^*_{Kep}$ to the separation distance $r$, which is the equilibrium observable semimajor axis. For the opposite perturbation, we instead increase the observable semimajor axis to be equal to the Keplerian semimajor axis. In other words we set 
\begin{equation}
    \dot{\theta}=\sqrt{\frac{\mu}{r^2}\left(\frac{2}{r}-\frac{1}{2a^*_{\text{Kep}}-r}\right)}.
\end{equation}
This corresponds to a perturbation of 
\begin{equation}
    \Delta h=r^2\sqrt{\frac{\mu}{r^2}\left(\frac{2}{r}-\frac{1}{2a^*_{\text{Kep}}-r}\right)}-h^*.
    \label{analytic_perturb2}
\end{equation}

\begin{figure}[ht!]
   \centering
   \includegraphics[width = 3in]{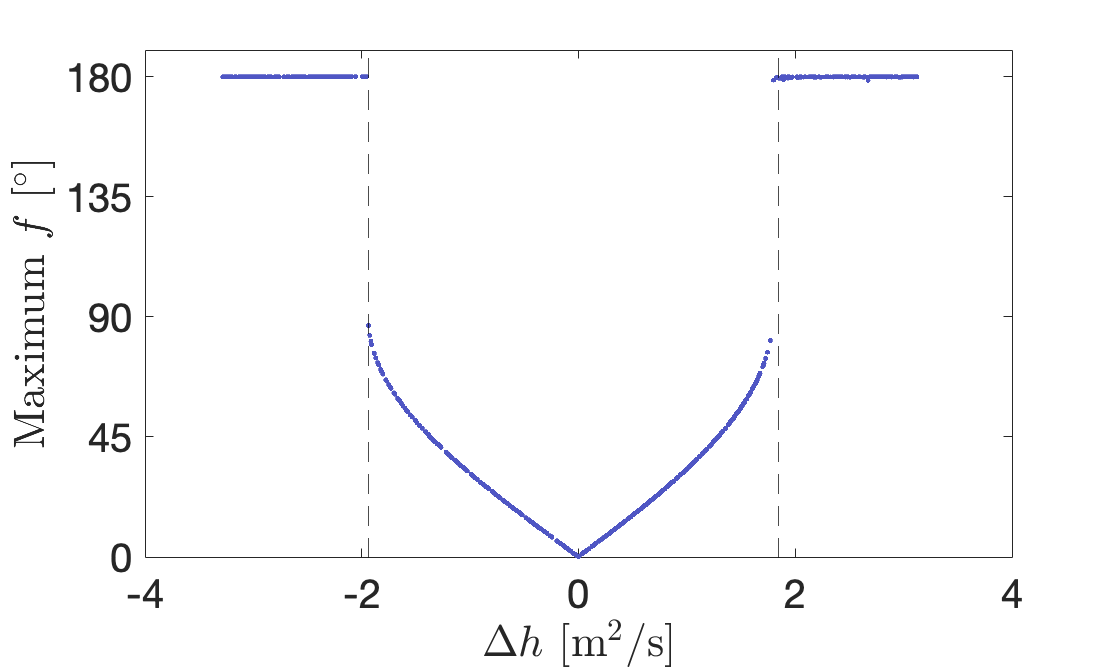} 
   \caption{The maximum true anomaly as a function of the change in specific orbital angular momentum for an orbital perturbation. The discontinuities correspond to perturbations sufficient to push the orbit out of equilibrium and allow the true anomaly to track the secondary. The vertical dashed lines correspond to the analytical solutions from Eq.\ \ref{analytic_perturb1} (left, indicating a decrease in the orbit period) and Eq.\ \ref{analytic_perturb2} (right, indicating an increase in the orbit period). Numerical simulations are plotted as individual points, and show strong agreement with the analytical formulae.}
   \label{deltah_full}
\end{figure}

\begin{figure}[ht!]
   \centering
   \includegraphics[width = 3in]{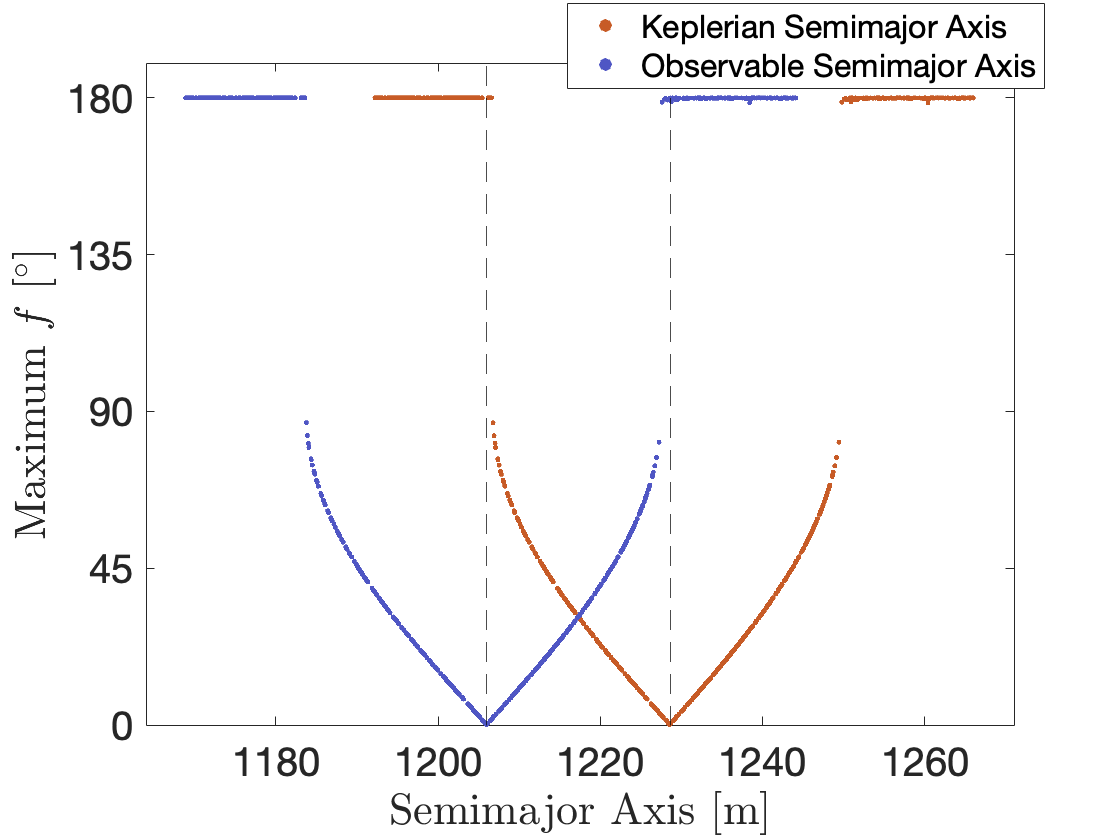} 
   \caption{The maximum true anomaly as a function of the change in both the observable and Keplerian semimajor axis. This illustrates how pushing a system out of equilibrium is equivalent to changing the semimajor axis by the difference between the observable and Keplerian values.}
   \label{deltah_sma}
\end{figure}

Fig.\ \ref{deltah_full} shows the full range of perturbations, both increasing and decreasing the orbit period. Numerical simulations calculated with \textsc{gubas} are plotted as data points along with the analytical solutions from Eqs.\ \ref{analytic_perturb1} and \ref{analytic_perturb2} as dashed lines. We see good agreement between these analytical solutions and the numerical results. Thus, we have established novel analytical equations to calculate the critical perturbation to the orbit necessary to break out of an equilibrium configuration and allow the true anomaly to track the secondary rather than librate about $0^\circ$. This perturbation is equivalent to changing the binary semimajor axis by $\pm(a^*_{\text{Kep}}-r)$, which is illustrated in Fig.\ \ref{deltah_sma}. 

\subsection{Lagrange Planetary Equations}
In their analysis of the binary asteroid (66391) Moshup, \cite{fahnestock2008simulation} developed Lagrange Planetary Equations (LPEs) for a binary system using a second-degree mutual potential. LPEs can provide a simple alternative to the full \textsc{gubas} integration. These equations have the advantage of an analytic method to calculate quantities such as the orbit precession without relying on more expensive numerical simulations. We do not reproduce the full set of LPEs here, but refer the reader to Eqs.\ 44--48 in \cite{fahnestock2008simulation}. Because the Didymos mutual orbit is assumed planar (i.e., the orbit angular momentum is aligned with the total angular momentum), we use the longitude of periapsis rather than the argument of periapsis and the longitude of the ascending node to avoid the singularity in the longitude of the ascending node. The longitude of periapsis is defined as the sum of these two classical Keplerian angles ($\bar{\omega}=\omega+\Omega$). For a planar system, the relevant LPEs simplify to:

\begin{equation}
    \dot{\bar{a}}_{\text{Kep}}=0
\end{equation}

\begin{equation}
    \dot{\bar{e}}_{\text{Kep}}=0
\end{equation}

\begin{equation}
    \dot{\bar{\omega}}=\frac{3\sqrt{\mu}}{2a_{\text{Kep}}^{7/2}(1-e_{\text{Kep}}^2)^2}\left(\bar{I}_{A,z}-\bar{I}_{A,xy}-2\bar{I}_{B,x}+\bar{I}_{B,y}+\bar{I}_{B,z}\right)
    \label{LPEprecession}
\end{equation}

\begin{equation}
    \dot{\bar{M}}=\sqrt{\frac{\mu}{a_{\text{Kep}}^3}}+\frac{3\sqrt{\mu}}{2a_{\text{Kep}}^{7/2}(1-e_{\text{Kep}}^2)^2}\left(\bar{I}_{A,z}-\bar{I}_{A,xy}-2\bar{I}_{B,x}+\bar{I}_{B,y}+\bar{I}_{B,z}\right).
\end{equation}

\begin{figure}[ht!]
   \centering
   \includegraphics[width = 5in]{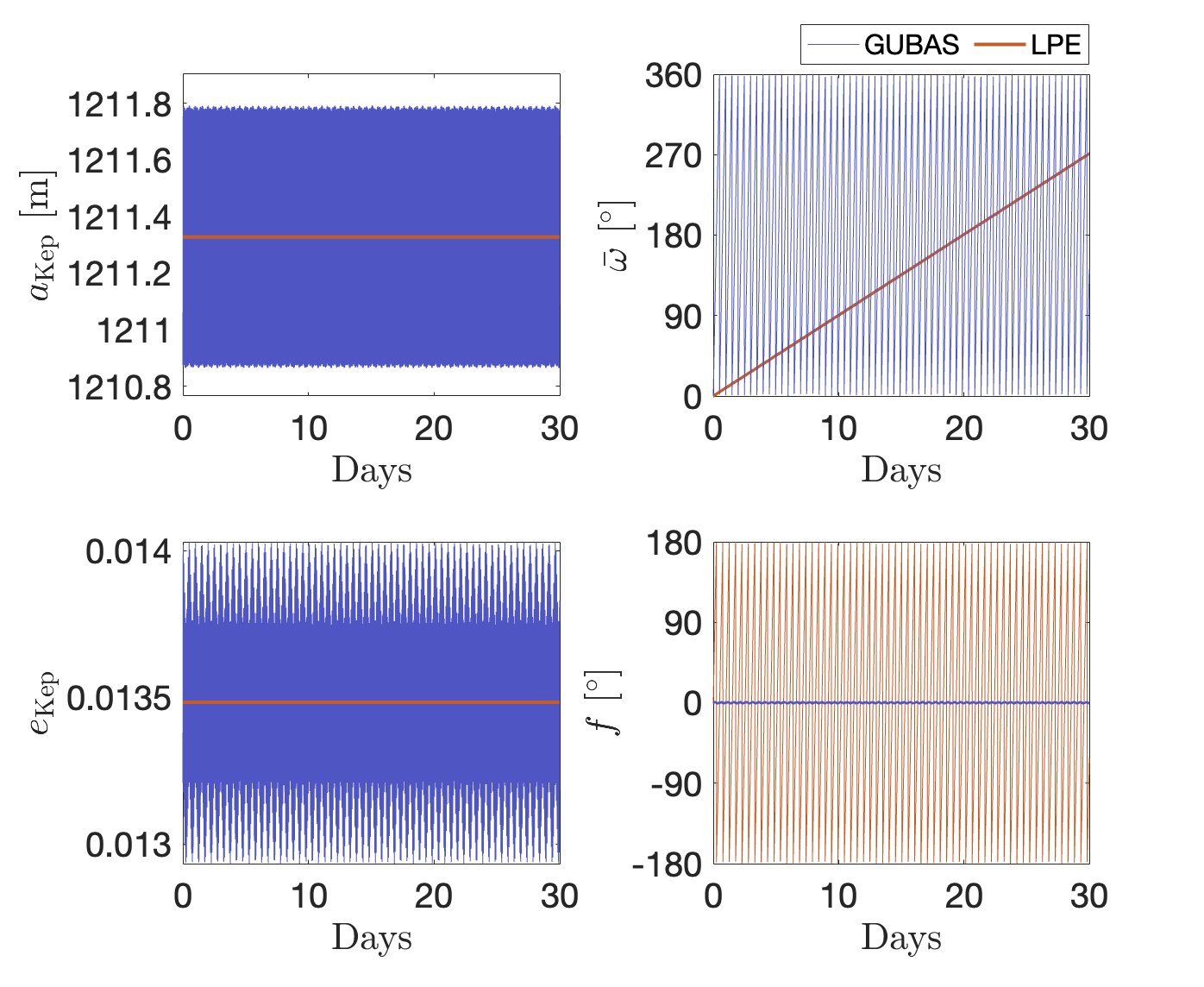} 
   \caption{The semimajor axis (top left), eccentricity (bottom left), longitude of periapsis (top right), and true anomaly (bottom right) of a numerical \textsc{gubas} simulation compared to the analytical LPE solution for an equilibrated system. The LPE solution shows the true anomaly circulating, while in reality this angle should be zero and the longitude of periapsis should be circulating.}
   \label{LPE_relaxed}
\end{figure}

First, we apply the averaged LPEs to an equilibrated Didymos system and compare to numerical simulations: see Fig.\ \ref{LPE_relaxed}. While there is agreement for the Keplerian semimajor axis and eccentricity, which remain constant, the LPEs do not correctly capture the average behavior of the geometric angles. Specifically, in the LPE solution the longitude of periapsis precesses and the true anomaly circulates, while in reality the true anomaly librates and the longitude of periapsis circulates. Thus, the LPEs are a poor representation of the dynamics when the true anomaly is librating.

\begin{figure}[ht!]
   \centering
   \includegraphics[width = 5in]{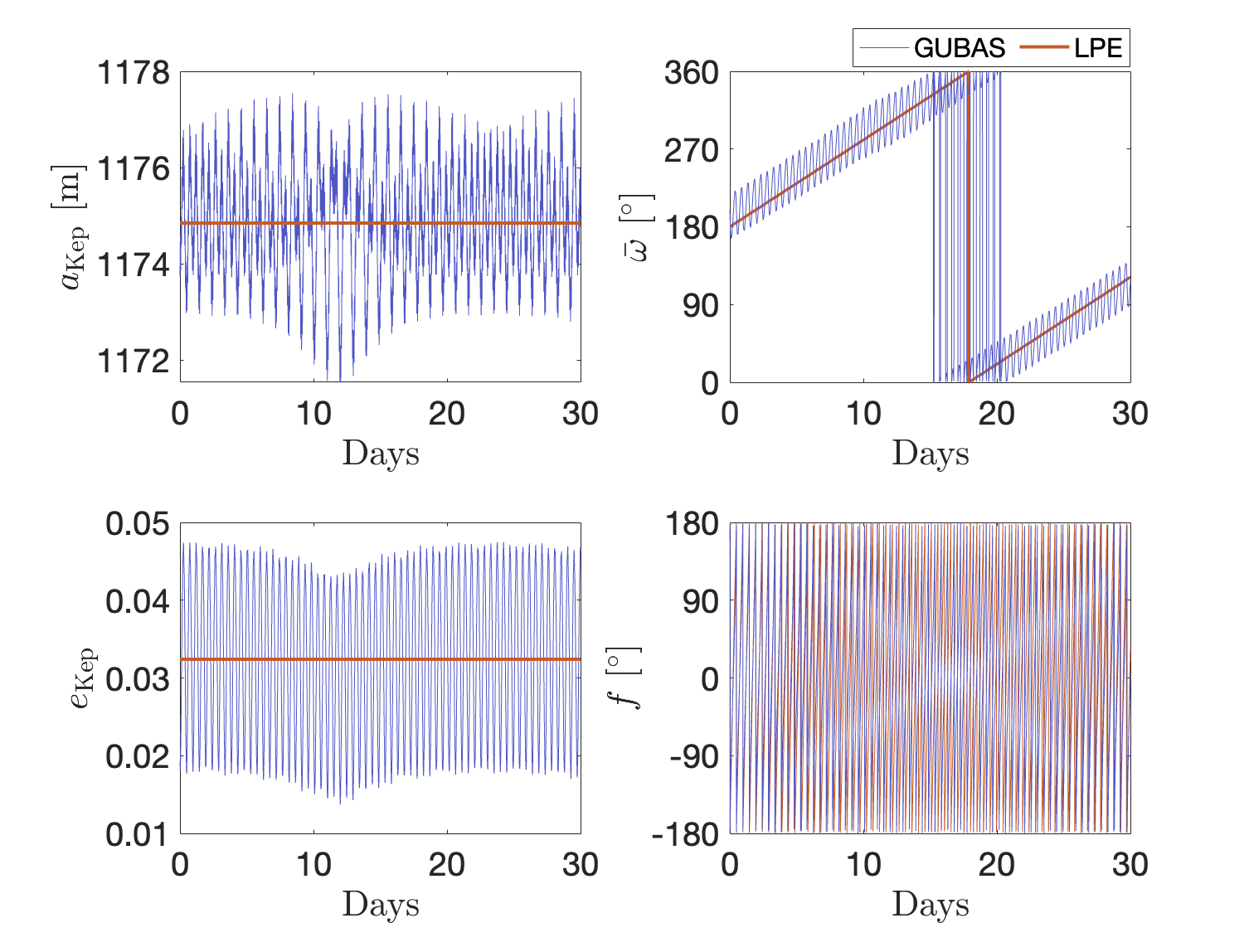} 
   \caption{The semimajor axis (top left), eccentricity (bottom left), longitude of periapsis (top right), and true anomaly (bottom right) of a numerical \textsc{gubas} simulation compared to the analytical LPE solution for a perturbed system. Because the system is perturbed, both \textsc{gubas} and the LPE correctly show the true anomaly circulating and the longitude of periapsis precessing.}
   \label{LPE_perturb}
\end{figure}

Compare this to an excited system, where there is better agreement between the LPEs and \textsc{gubas} solutions, as seen in Fig.\ \ref{LPE_perturb}. While the LPEs correctly compute the average values of semimajor axis and eccentricity, they do not capture the fluctuations that these variables experience. In the perturbed system, the LPEs correctly calculate the behavior of the longitude of periapsis, where this angle precesses. For the true anomaly, again the LPEs correctly show this angle circulating, although with a period slightly different from the numerical solution. Nonetheless, the LPEs are an effective approach for quickly calculating the precession period without the need for expensive simulations, as long as the system is not in an equilibrium state. One shortfall of the LPEs is they ignore any triaxiality of the secondary \citep{fahnestock2008simulation}. For an oblate spheroidal shape like Dimorphos, as derived from DRACO images \citep{daly2023}, this is a fine assumption, but for more elongated secondaries experiencing libration, the LPEs will incorrectly calculate the precession rate.

The cause of the LPEs' shortfalls for the relaxed system stems from their definition. The LPEs are calculated by averaging over the mean anomaly. However, this is not appropriate for a system in equilibrium, as the mean anomaly remains equal to zero. Rather than using LPEs for a system in equilibrium, one can instead simply use Eq.\ \ref{f2bp_eq} in place of $\dot{\bar{\omega}}.$ Once a system is perturbed out of the equilibrium and the mean anomaly is allowed to circulate, the LPEs become more accurate. However, for a significantly triaxial secondary experiencing libration, a higher-fidelity model should be used, for example the analytic correction defined in \cite{borderies1990phobos} or the semi-analytic model in \cite{cuk2010orbital}. Given the limited scope of applicability of the averaged LPEs, we caution their use in the analysis of binary asteroid dynamics.

\section{Effects of the DART impact} \label{sec:dart}
We now focus on the actual DART impact and how it changed the Didymos system mutual orbit. We can calculate the change in specific orbital angular momentum to predict if the true anomaly should be librating or circulating. Assuming a circular pre-impact orbit and a planar, head-on impact with the secondary, the impulsive change in specific orbital angular momentum is easily calculated:

\begin{equation}
    \Delta h = r\Delta v_T.
\end{equation}
For $\Delta v_T = -2.7$ mm/s \citep{cheng2023}, the change in specific orbit angular momentum is about $-$3.2 m$^2$/s. While the actual impact conditions are more complicated than this simple analysis, this value is sufficiently beyond the circulation threshold of roughly $-$2 m$^2$/s from Fig.\ \ref{deltah_decrease} and Eq.\ \ref{analytic_perturb1} that we can confidently predict that the DART impact changed the system's momentum sufficiently to push it out of an equilibrium state. Furthermore, since the DART impact provided a sufficient perturbation to break the equilibrium configuration, the LPEs can provide an accurate way to calculate the apsidal precession rate of the post-impact orbit. Under a uniform density assumption, the apsidal precession rate of the orbit varies roughly between $8-20^\circ$/day considering the system uncertainties, with a large dependence on the $J_2$ gravity coefficient of Didymos. With the range of plausible Didymos shapes from Table \ref{parameters}, its $J_2$ coefficient ranges from about 0.07 to 0.11, again under a uniform mass density assumption. Mass heterogeneities allow for a wider range of possible values for Didymos's $J_2$ and the mutual orbit apsidal precession rate.

To estimate the changes in Dimorphos's orbit, we use \textsc{gubas} in Monte Carlo simulations to model the system dynamics. We follow the approach from \cite{cheng2023}, randomly drawing from the independent uncertainties in the system parameters listed in Table \ref{parameters}. Each Monte Carlo realization uses numerical routines to draw one set of possible parameters from the distributions listed in Table \ref{parameters}, then uses the randomly selected pre-impact orbit period to calculate the system's mass for an initially (physically) circular orbit. The algorithm then iterates the change in velocity ($\Delta \vec{v}$) to match the randomly selected post-impact orbit period. We also apply a random radial and out-of-plane component, equal to anywhere between 0 and 50\% of the in-plane $\Delta v_T$, to account for the uncertainty in the full three-dimensional $\Delta \vec{v}$ caused by the ejecta cone. This is equivalent to the full three-dimensional $\Delta \vec{v}$ vector being anywhere within $30^\circ$ of the orbit tangent direction, consistent with estimates from \cite{cheng2023}. 

For our numerical Monte Carlo analysis of the post-impact dynamics, we first outline our approach to calculating the system mass. We then calculate the change in velocity caused by the DART impact. Next, we show the resulting change in mutual orbit semimajor axis. Lastly, we discuss the change in eccentricity.

\subsection{Mass Calculation}
\begin{figure}[ht!]
   \centering
   \includegraphics[width = 3in]{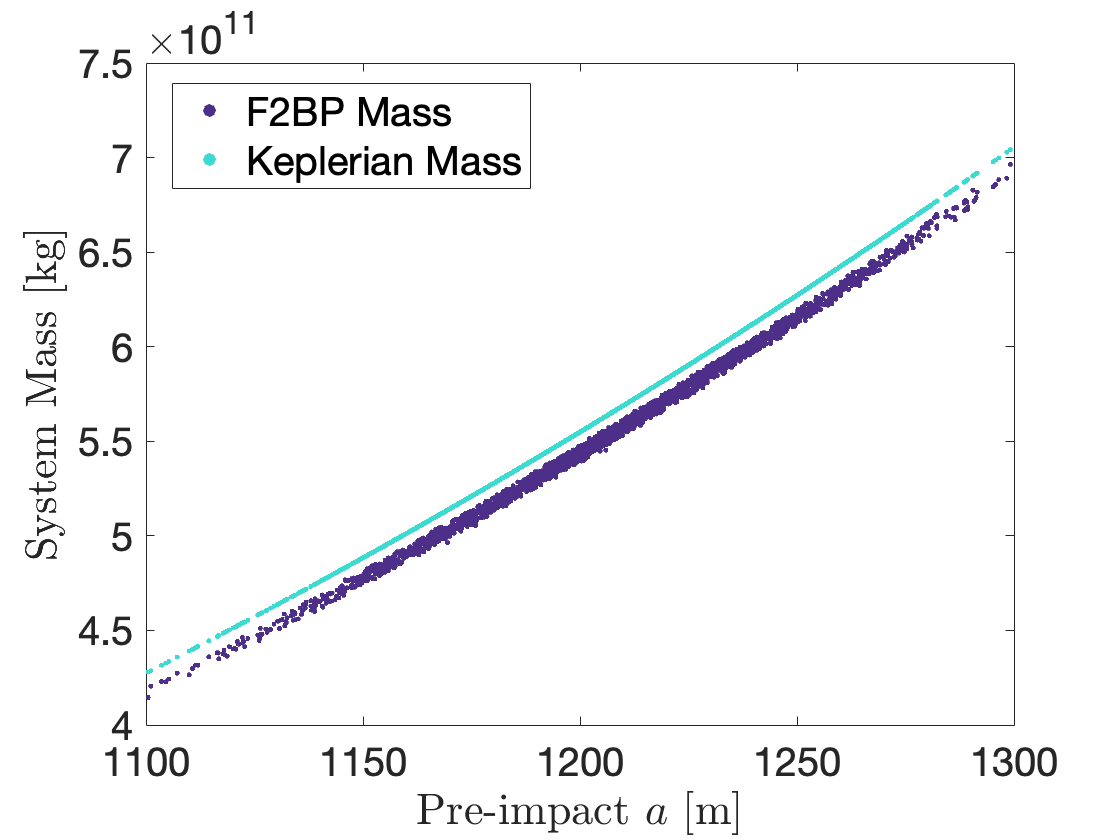} 
   \caption{System mass calculated using a numerical routine considering F2BP dynamics, along with the system mass calculated using Keplerian dynamics, as functions of the pre-impact observable semimajor axis. Ignoring the F2BP coupling and Didymos oblateness results in a consistently larger mass estimate.}
   \label{f2bp_mass}
\end{figure}

An important advantage inherent to binary asteroids is the observability of the system's mass through measurements of the mutual orbit period. Unfortunately, due to spin-orbit coupling, the classical Keplerian approach is not applicable for a close, irregularly shaped binary like the Didymos system. Instead, we use a numerical secant search algorithm similar to that used in \cite{agrusa2021excited}. In the secant search, the system bulk density at iteration $n$ is

\begin{equation}
\rho_n = \rho_{n-1}-\Delta T(\rho_{n-1})\frac{\rho_{n-1}-\rho_{n-2}}{\Delta T(\rho_{n-1})-\Delta T(\rho_{n-2})},
\label{rho_secant}
\end{equation}
where the error function $\Delta T$ is simply the difference between the measured orbit period and the average stroboscopic orbit period from simulations. Starting from the Keplerian mass for the initial guess, we iterate the solution until this error function is smaller than the uncertainty of the orbit period. This is the approach used in the Monte Carlo simulations for the mass calculation considering the F2BP dynamics. The mass results are plotted as a function of the pre-impact observable semimajor axis in Fig.\ \ref{f2bp_mass}. This is also compared to the mass calculated using purely two-body Keplerian dynamics. The Keplerian mass is systematically larger than the true F2BP mass, primarily due to the oblateness of Didymos.

\begin{figure}[ht!]
   \centering
   \includegraphics[width = 3in]{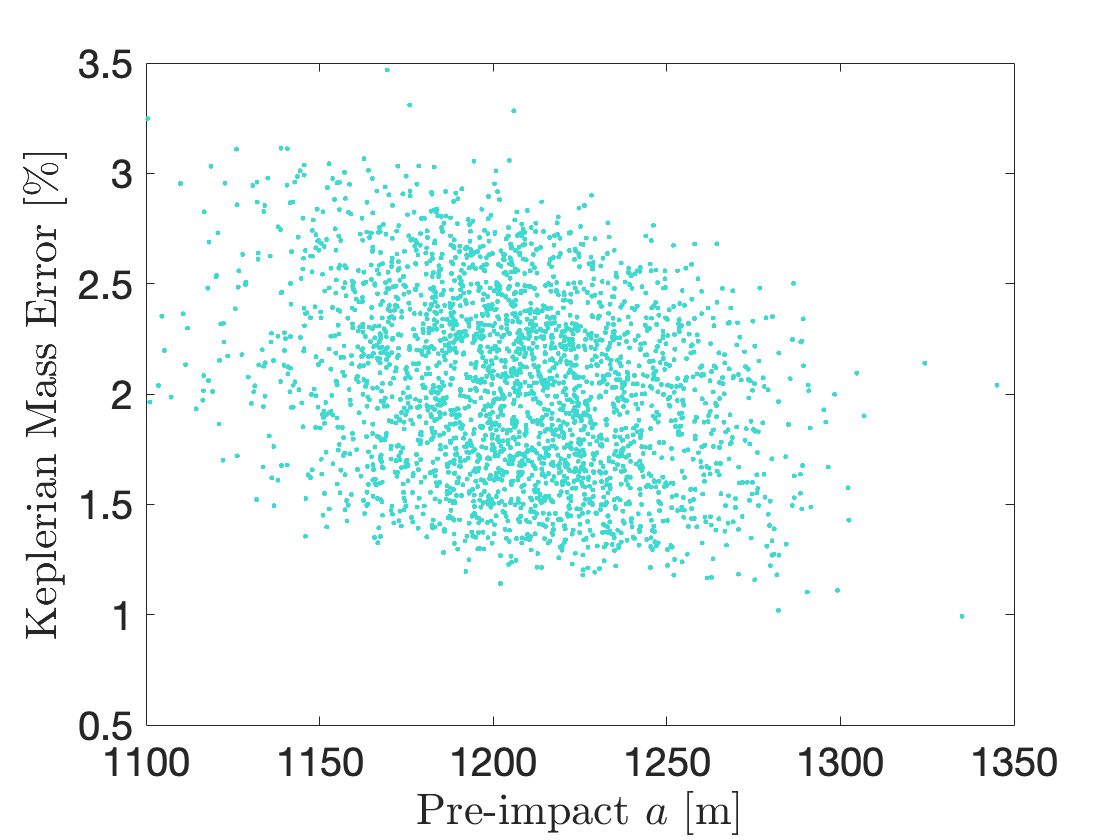} 
   \caption{Percent error in the calculated mass when ignoring the aspherical shapes and F2BP dynamics, as a function of pre-impact observable semimajor axis. Using Keplerian dynamics results in approximately a 1--3\% error in mass estimate.}
   \label{f2bp_mass_error}
\end{figure}

The error from using the Keplerian mass is shown in Fig.\ \ref{f2bp_mass_error}, which ranges from around 1 to 3\% relative to the ``true'' mass calculated from the F2BP dynamics. This highlights the need to consider the aspherical shapes and spin-orbit coupling in estimating the mass of close, irregular-shaped binary asteroids. However, as a rough first-order estimate for Didymos, the true system mass is around $98\%$ the calculated Keplerian mass.

\begin{figure}[ht!]
   \centering
   \includegraphics[width = 3in]{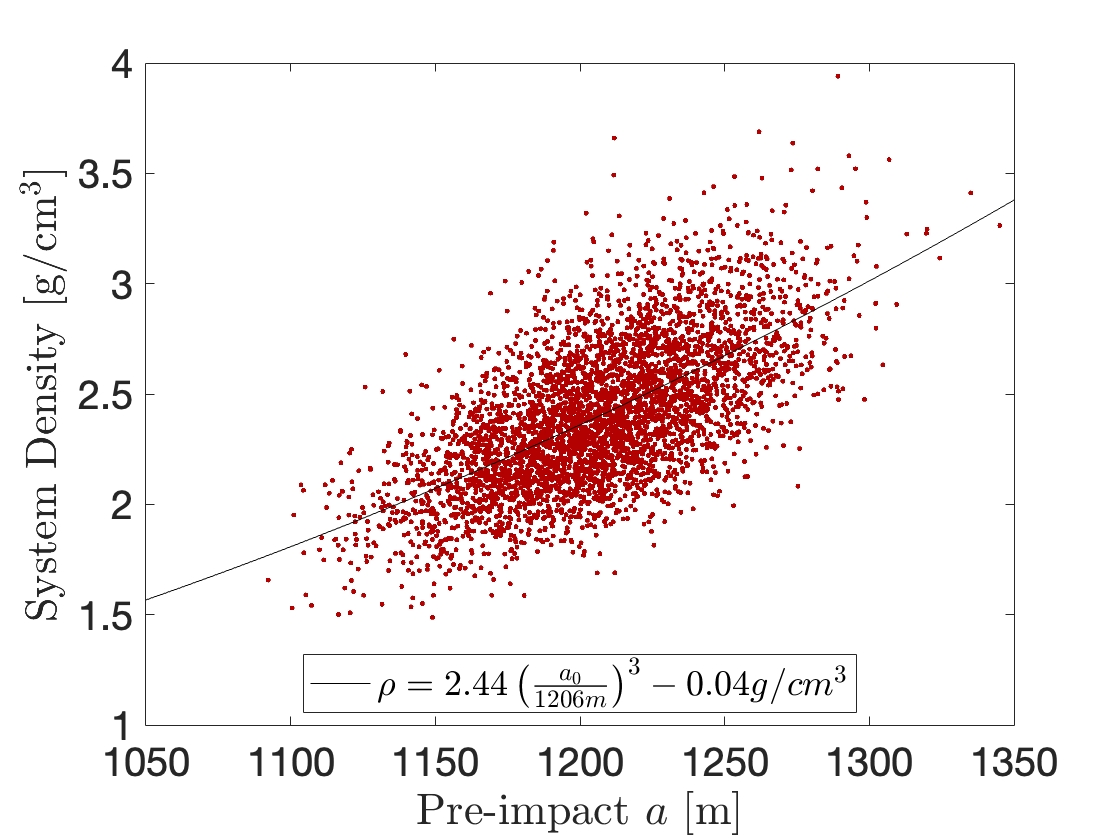} 
   \caption{System bulk density as a function of the possible pre-impact observable semimajor axis, calculated using a numerical routine considering F2BP dynamics. Accounting for the system uncertainties, including the volumes, the bulk density is $2.4\pm0.3$ g/cm$^3$ ($1\sigma$), with a dependence on the pre-impact separation.}
   \label{f2bp_density}
\end{figure}

As a convenient quantity for comparison with other asteroids, we calculate the bulk density of the system. This is shown as a function of the pre-impact observable semimajor axis in Fig.\ \ref{f2bp_density}. Note this is the density the system would have if the primary and secondary had equal, homogeneous densities. The bulk density is about $2.4\pm0.3$ g/cm$^3$ ($1\sigma$), which matches results from \cite{daly2023}, but we highlight the dependency on the pre-impact observable semimajor axis. Based on the Keplerian relationship, we fit the bulk density as a cubic function of the semimajor axis: $\rho \approx 2.44\left(\frac{a_0}{1206\text{ m}}\right)^3-0.04\text{ g/cm}^3$, where $a_0$ is the pre-impact observable semimajor axis in meters. However, there is considerable uncertainty around this line, largely due to the bodies' volume uncertainties, and this model should only be used as a rough approximation. Furthermore, the fit only applies to ranges of pre-impact observable semimajor axis used here and should not be extrapolated.

\subsection{Change in Velocity}
\begin{figure}[ht!]
   \centering
   \includegraphics[width = 3in]{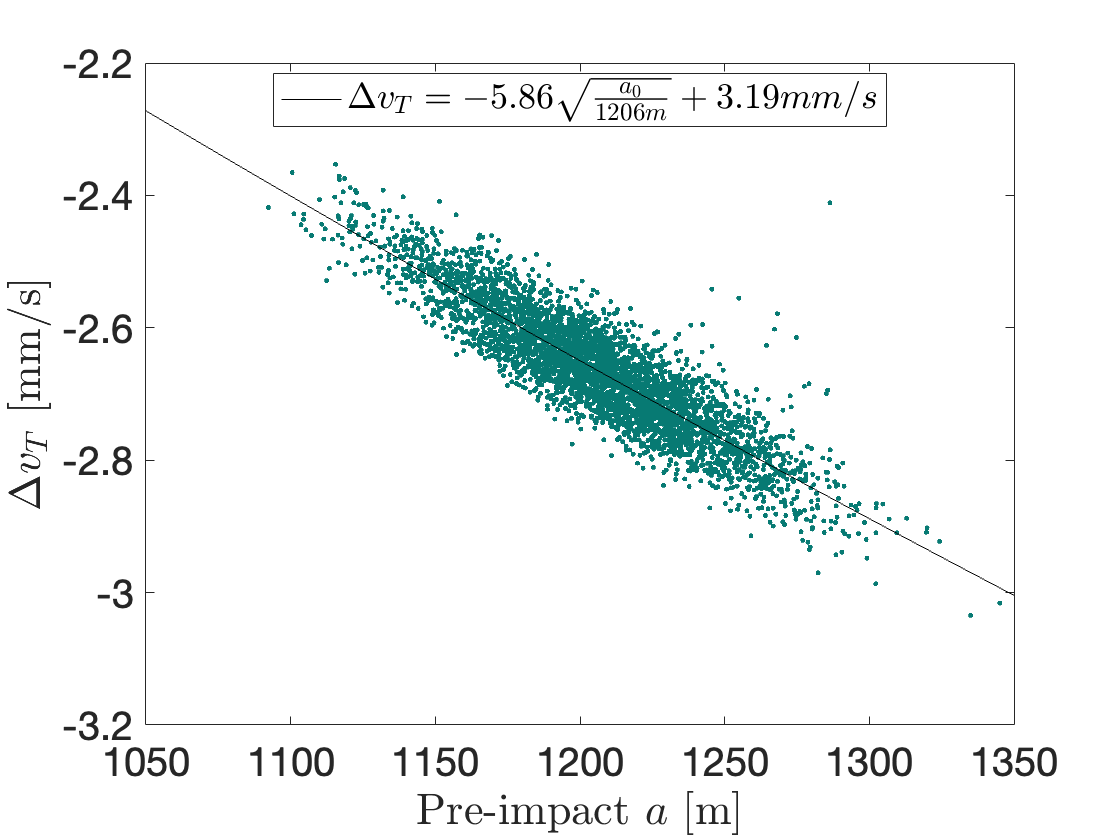} 
   \caption{Change in along-track velocity as a function of the possible pre-impact observable semimajor axis. There is a strong relationship between $\Delta v_T$ and pre-impact observable semimajor axis.}
   \label{delta_v}
\end{figure}

Next, we calculate the along-track $\Delta v_T$ necessary to achieve the observed post-impact orbit period. While we include three-dimensional components in our analysis, we find essentially only the along-track change in velocity has an effect on the orbit period. As a function of the pre-impact observable semimajor axis, the along-track $\Delta v_T$ is plotted in Fig.\ \ref{delta_v}. Based on the Keplerian relationship, we fit $\Delta v_T$ as a function of the square-root of the pre-impact semimajor axis: $\Delta v_T \approx -5.86\sqrt{\frac{a_0}{1206\text{ m}}}+3.19\text{ mm/s}$, where again $a_0$ is the pre-impact observable semimajor axis in meters. Again, this numerical fit only applies to the range of data shown in Fig.\ \ref{delta_v} and should not be extrapolated out of this domain. Our results agree with those presented in \cite{cheng2023}, with a $\Delta v_T=-2.7\pm0.1$ mm/s, however here we highlight the dependence on the pre-impact observable semimajor axis. This also demonstrates that any component of the full $\Delta v$ vector not aligned with the orbit tangent direction does not affect the post-impact orbit period and is thus unobservable from the ground.

\begin{figure}[ht!]
   \centering
   \includegraphics[width = 3in]{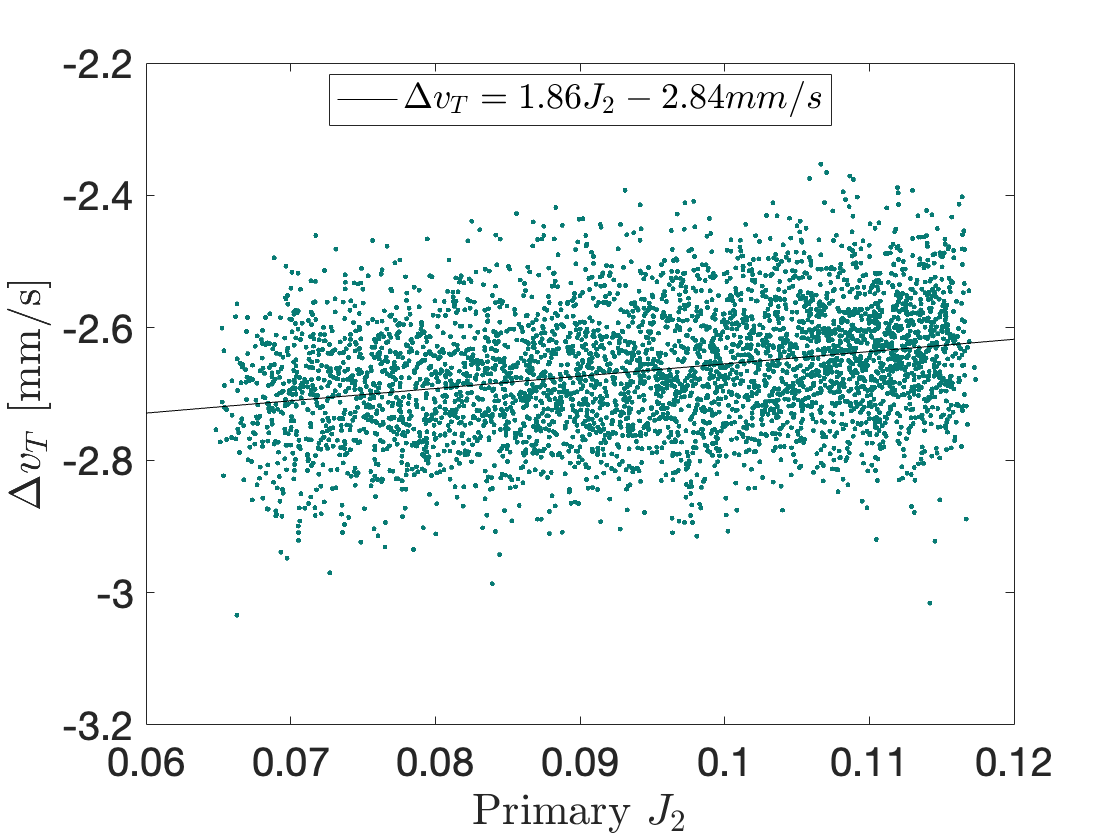} 
   \caption{Change in along-track velocity as a function of Didymos's possible $J_2$ gravity coefficient. The magnitude of $\Delta v_T$ decreases slightly as Didymos's $J_2$ increases.}
   \label{delta_v_J2}
\end{figure}

Besides the dependence of $\Delta v_T$ on the pre-impact observable semimajor axis, we also find a smaller dependence on Didymos's oblateness, or its $J_2$ gravity term. This is shown in Fig.\ \ref{delta_v_J2}, where as Didymos becomes more oblate (larger $J_2$ values), the magnitude of $\Delta v_T$ needed to achieve the post-impact orbit period decreases. While the dependence of $\Delta v_T$ on the pre-impact semimajor axis dominates, it's also important to estimate Didymos's $J_2$ coefficient to fully understand the effects of the DART impact.

\subsection{Change in Semimajor Axis}
\begin{figure}[ht!]
   \centering
   \includegraphics[width = 3in]{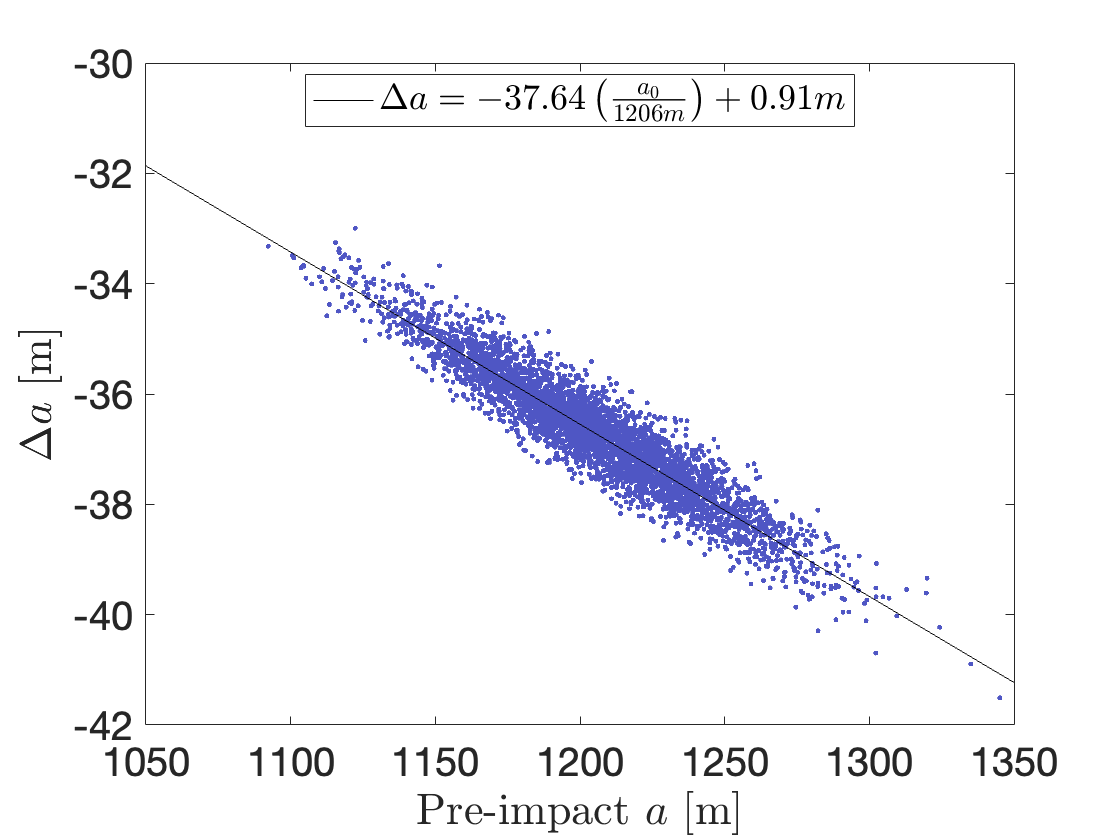} 
   \caption{The change in observable semimajor axis as a function of the possible pre-impact observable semimajor axis. There is a strong linear relationship between the post- and pre-impact semimajor axes.}
   \label{delta_a}
\end{figure}

While \cite{cheng2023} calculated the change in velocity caused by the impact, they do not discuss the resultant changes to the orbit. Here we calculate the change in the observable semimajor axis. From the Monte Carlo simulations we calculate the post-impact observable semimajor axis and solve for the change in the observable semimajor axis. Fig.\ \ref{delta_a} plots the change in observable semimajor axis as a function of the pre-impact observable semimajor axis. We find $\Delta a=-37\pm1$ m ($1\sigma$), but the change in semimajor axis depends on the pre-impact observable semimajor axis. Again, using a simple linear fit, we find $\Delta a \approx -37.64\left(\frac{a_0}{1206\text{ m}}\right)+0.91\text{ m}$. From the Monte Carlo simulations, the post-impact observable semimajor axis is also directly calculated, equal to $1170\pm34$ m ($1\sigma$).

Thanks to the linear relationship between the pre- and post-impact observable semimajor axis, it is simple for measurements of the post-impact semimajor axis to add additional constraints to the pre-impact value. This is important, as we have already seen the importance of tightening the constraints on the pre-impact observable semimajor axis, as $\Delta v_T$ strongly depends on this quantity. Thus, a major advantage and contribution of ESA's Hera mission, planned to return to Didymos to study the post-impact system in early 2027, is the ability to precisely measure the current semimajor axis \citep{michel2022esa}.

This linear relationship is expected from basic Keplerian dynamics. From Kepler's third law, we know the following relationship between orbital period and semimajor axis for a common barycenter:
\begin{equation}
    \frac{P_1^2}{a_1^3} = \frac{P_2^2}{a_2^3}.
\end{equation}
This can be rearranged and, substituting in the parameters from Table \ref{parameters}, we find
\begin{equation}
    a_2 = a_1\left(\frac{P_2}{P_1}\right)^{2/3} = 1168.65\left(\frac{a_1}{1206 \text{ m}}\right) \text{ m}.
\end{equation}
Even though the F2BP dynamics are non-Keplerian in reality, applying Kepler's third law provides a relatively accurate estimate for the relationship between the pre- and post-impact semimajor axis. Furthermore, this illustrates the linear relationship. Thus, precise measurements from Hera of the post-impact semimajor axis also provide estimates for the pre-impact semimajor axis. This in turn leads to a more accurate estimate for the system's mass and the $\Delta v_T$ of the DART impact, provided the orbit has only experienced minimal secular evolution since the impact.

\subsection{Change in Eccentricity}
\begin{figure}[ht!]
   \centering
   \includegraphics[width = 3in]{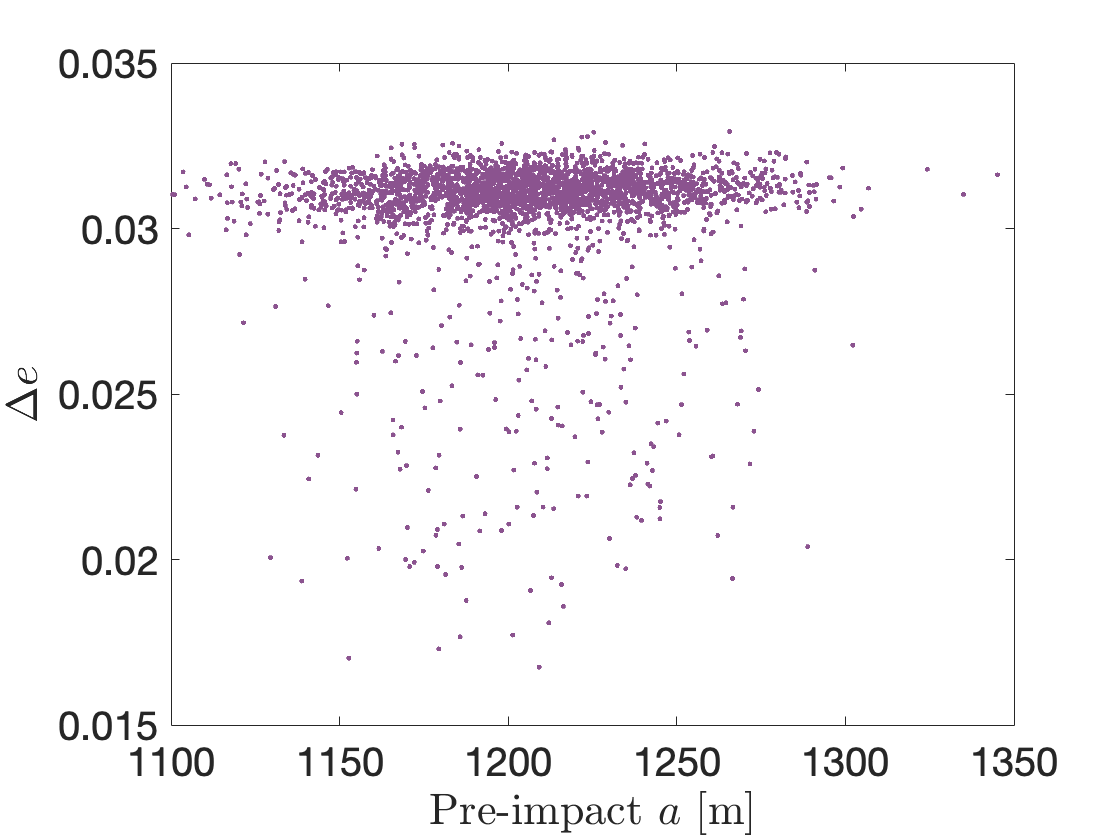} 
   \caption{Change in observable eccentricity as a function of the possible pre-impact observable semimajor axis. The change in eccentricity is largely independent of the pre-impact semimajor axis. The pre-impact orbit is assumed physically circular. The simulation runs with a skewed smaller eccentricity change are the result of tumbling in the secondary.}
   \label{delta_e}
\end{figure}

Next, we discuss the post-impact eccentricity, which has also not yet been included in the literature. The change in observable eccentricity is plotted in Fig.\ \ref{delta_e} as a function of the pre-impact observable semimajor axis. Notably, while previous quantities have depended strongly on the pre-impact observable semimajor axis, the eccentricity is largely independent of this quantity. From an initial orbit with zero observable eccentricity, the post-impact eccentricity is $0.031\pm0.002$.

It is worth discussing the skewed distribution in Fig.\ \ref{delta_e}, in which the change in observable eccentricity is occasionally much smaller than the representative value of 0.031. Each data point in Fig.\ \ref{delta_e} is the average post-impact observable eccentricity over the full simulation time of 100 days. The lower averages correspond to simulations in which the secondary enters a state of non-principal axis (NPA) rotation. Approximately $18\%$ of our simulation runs result in this attitude instability. This is an interesting dynamical state with major implications for observations of Didymos, and is the subject of Section \ref{sec:tumbling}.

\begin{figure}[ht!]
   \centering
   \includegraphics[width = 3in]{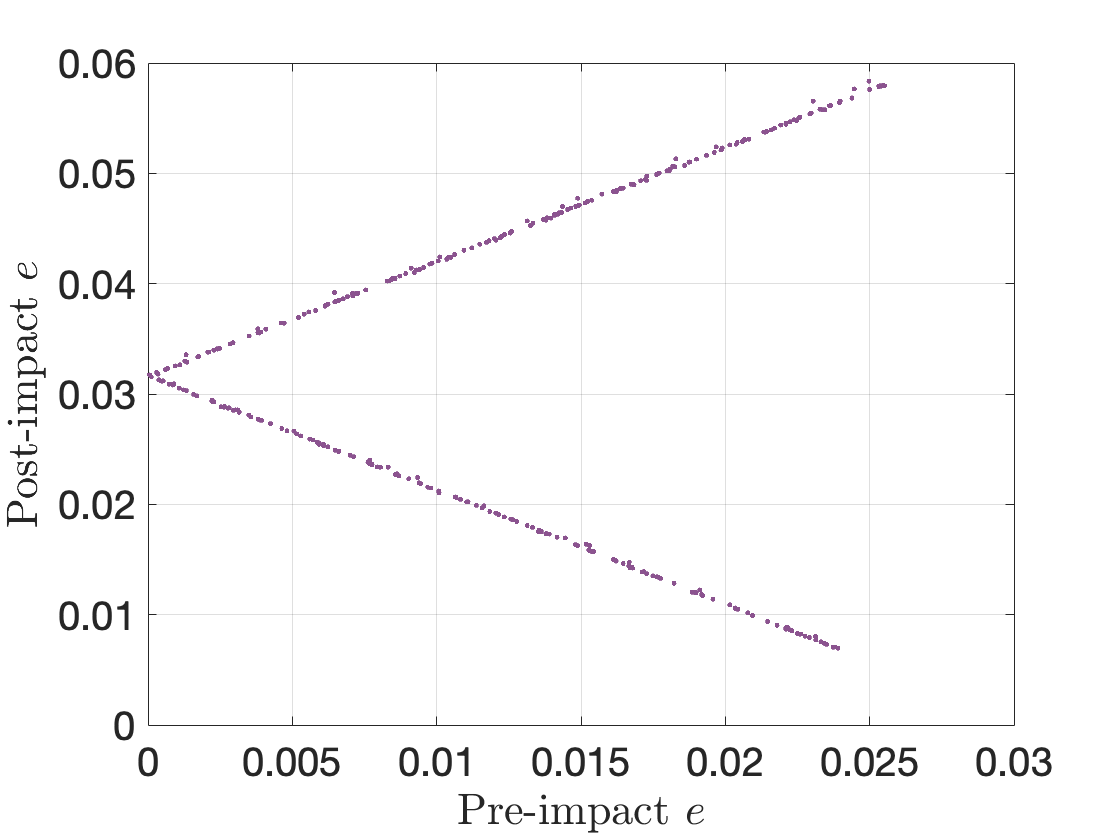} 
   \caption{The post-impact observable eccentricity as a function of the pre-impact observable eccentricity. The two curves define an envelope in which the post-impact observable eccentricity may exist, depending on where in its orbit the secondary was at the time of the perturbation.}
   \label{delta_e2}
\end{figure}

Next, we relax the initially circular orbit assumption. Using only the nominal system (i.e., the parameters from Table \ref{parameters} without the uncertainties), we vary the pre-impact observable eccentricity. Fig.\ \ref{delta_e2} shows the post-impact observable eccentricity as a function of the pre-impact observable eccentricity. In these simulations, we only apply a tangential $\Delta v_T$, and the perturbation occurs when the secondary is either at apoapsis or periapsis of the now-eccentric pre-impact orbit. This means the two curves in Fig.\ \ref{delta_e2} define an envelope of permissible eccentricity values. Depending on where in the orbit the secondary is at the time of the impact, the resulting eccentricity can lie anywhere in the area bounded by the two curves. Thus, measuring the post-impact eccentricity can give constraints to the pre-impact eccentricity. However, if the pre-impact orbit was eccentric, it is unknown where in its orbit Dimorphos was at the time of the impact, so a precise relationship between pre- and post-impact eccentricity is impossible.

\begin{figure}[ht!]
   \centering
   \includegraphics[width = 3in]{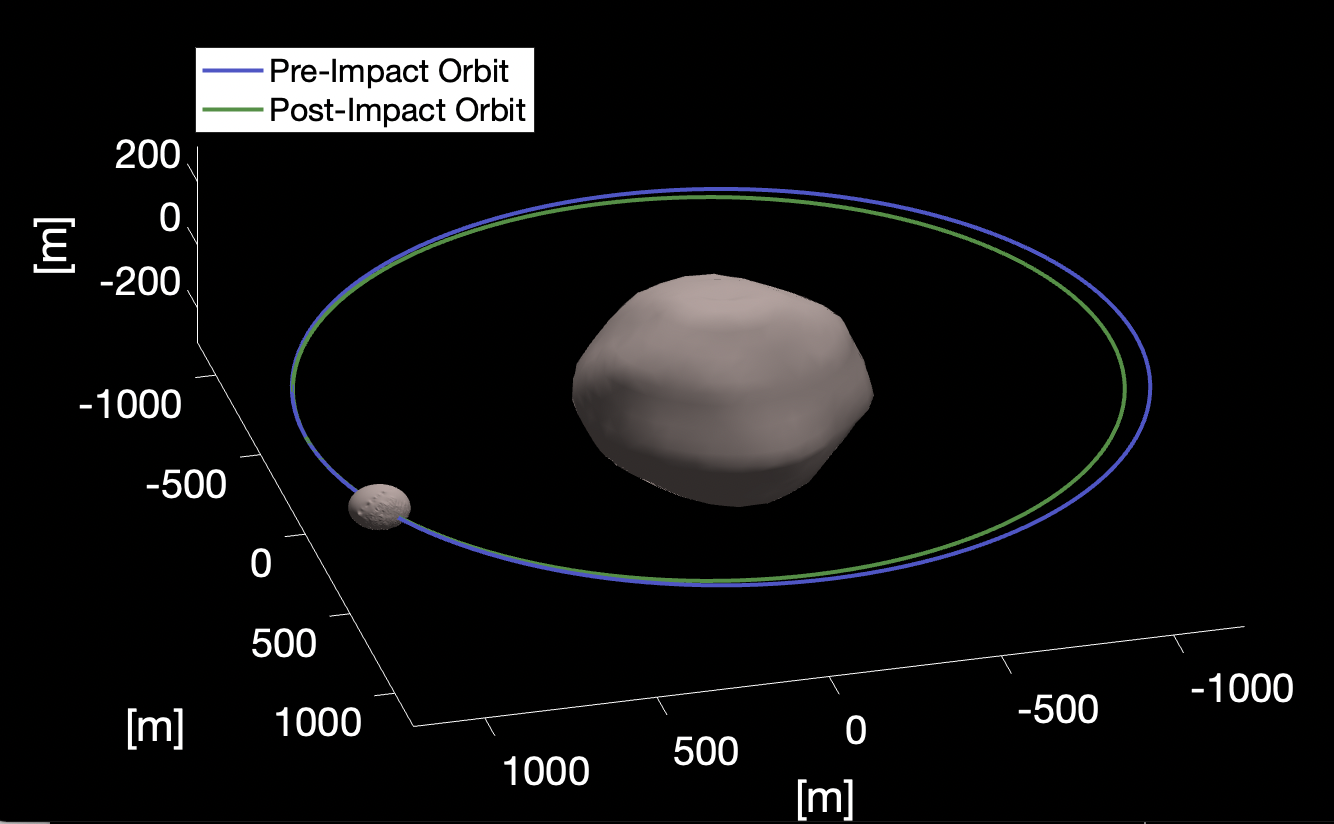} 
   \caption{The pre- and post-impact orbits for the nominal system plotted to scale. For illustration purposes, the scaled radar shape model is used to show Didymos \citep{naidu2020radar} and the \cite{daly2023} shape model is used to show Dimorphos.}
   \label{orbit}
\end{figure}

To illustrate the change in Dimorphos's orbit, we plot the system with the nominal pre- and post-impact orbits ignoring the uncertainties in Table \ref{parameters} in Fig.\ \ref{orbit}. This shows the differences in the mutual orbit caused by the DART impact. For illustration purposes we use the radar shape model from \cite{naidu2020radar} scaled to the size calculated by \cite{daly2023} for Didymos, and the DRACO shape model of Dimorphos from \cite{daly2023}. The outer orbit is the pre-impact, and the inner orbit is the post-impact. Only the first full orbit after the impact is plotted, and over time this orbit will precess around Didymos.

\section{Onset of Tumbling} \label{sec:tumbling}
In a perturbed orbit, it's possible for the secondary to begin tumbling due to resonances among the system's natural frequencies \citep{agrusa2021excited}. When this happens, the rotational angular momentum of the secondary will decrease on average as its spin axis moves away from its major principal inertia axis. As a result, the orbital angular momentum will on-average increase to compensate, which reduces the orbit's eccentricity. If the increase in orbital angular momentum, and corresponding decrease in eccentricity, is sufficient, it can cause the orbit to move back to an equilibrium configuration where the true anomaly begins librating again and the argument of periapsis tracks the secondary. Thus, unstable secondary rotation actually leads the orbit closer to a stable equilibrium. The nearest equilibrium orbit to the perturbed orbit is at the same value of observable semimajor axis but at the equilibrium eccentricity. Thus, we can simply apply Eq.\ \ref{equilibrium} calculated at the new observable semimajor axis. 

\begin{figure}[ht!]
   \centering
   \includegraphics[width = 3in]{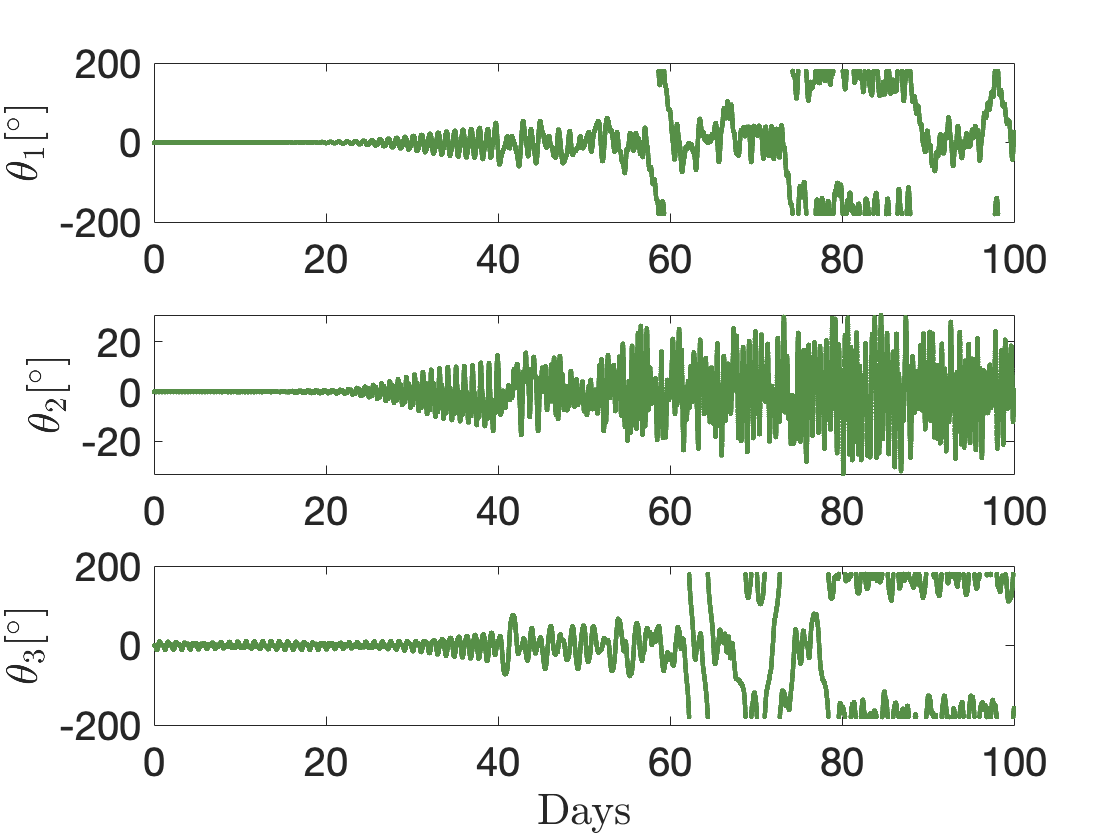} 
   \caption{The relative 1-2-3 Euler angles (roll-pitch-yaw) of the secondary of an unstable system that begins tumbling. Significant non-principal-axis rotation begins at around 40 days into the simulation.}
   \label{tumble_euler}
\end{figure}

To illustrate this point, we use an example of an unstable system calculated in Sec.\ \ref{sec:dart}. The relative 1-2-3 Euler angles, corresponding to roll-pitch-yaw of the secondary, are shown in Fig.\ \ref{tumble_euler}. This system demonstrates an attitude instability as the secondary has significant non-principal-axis rotation beginning around 40 days into the simulation. The Euler angles also reveal that, even while the secondary is tumbling, it still remains either generally aligned or anti-aligned with Didymos. It occasionally switches between states where its long axis is on-average pointing toward Didymos ($\theta_1$ and $\theta_3$ oscillating around $0^\circ$) to pointing away from Didymos ($\theta_1$ and $\theta_3$ oscillating around $180^\circ$).

\begin{figure}[ht!]
   \centering
   \includegraphics[width = 3in]{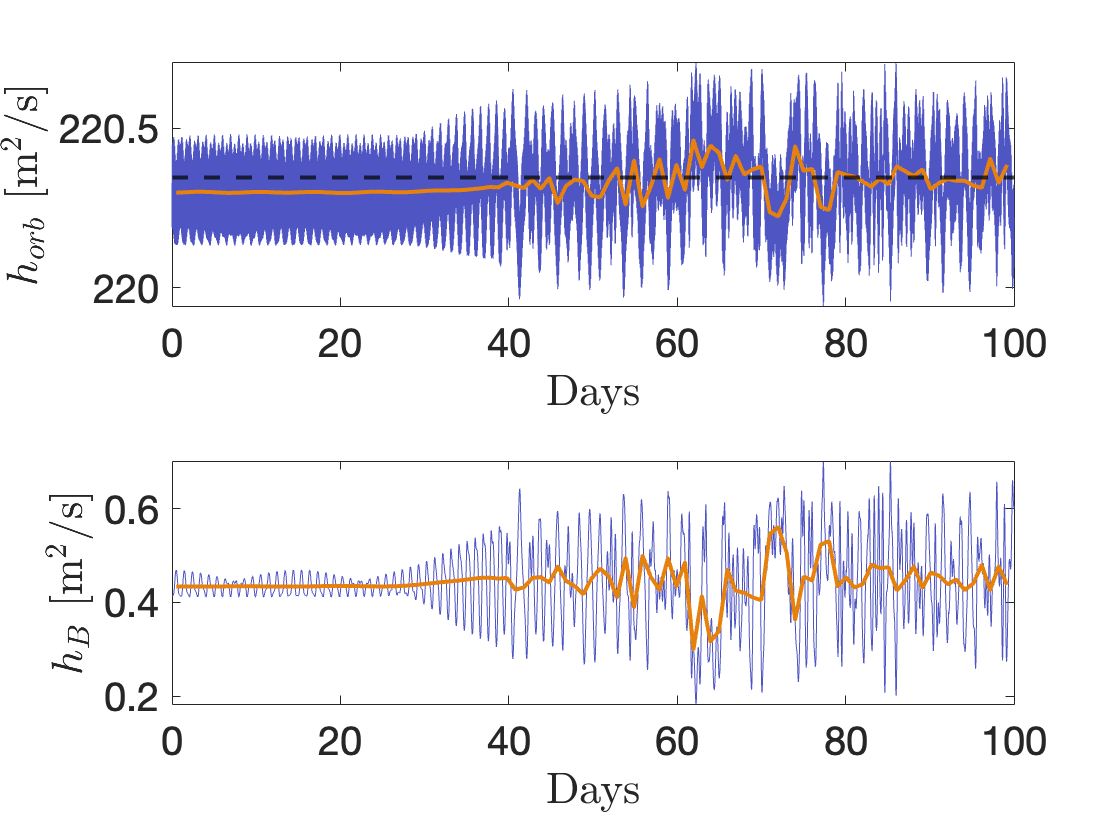} 
   \caption{The specific orbital angular momentum (top) and secondary angular momentum (bottom) of a system that begins tumbling. As the secondary enters a tumbling state, the orbital angular momentum increases on average. The dashed line corresponds to the nearest equilibrium orbit, and the orange line is a 1-day moving average of the angular momentum. We see an exchange of angular momentum between the orbit and the secondary when the secondary is tumbling.}
   \label{tumble_h}
\end{figure}

The specific orbital and secondary angular momenta of this system are plotted in Fig.\ \ref{tumble_h}. As the secondary enters a state of tumbling, the orbital angular momentum increases on average. The dashed black line in Fig.\ \ref{tumble_h} corresponds to the calculated angular momentum of the nearest equilibrium orbit from Eq.\ \ref{equilibrium} using the post-impact observable semimajor axis. The orange line is a 1-day average of the angular momentum to aid in interpretation. As the secondary tumbles, there is an exchange of angular momentum between the secondary and the primary. 

\begin{figure}[ht!]
   \centering
   \includegraphics[width = 3in]{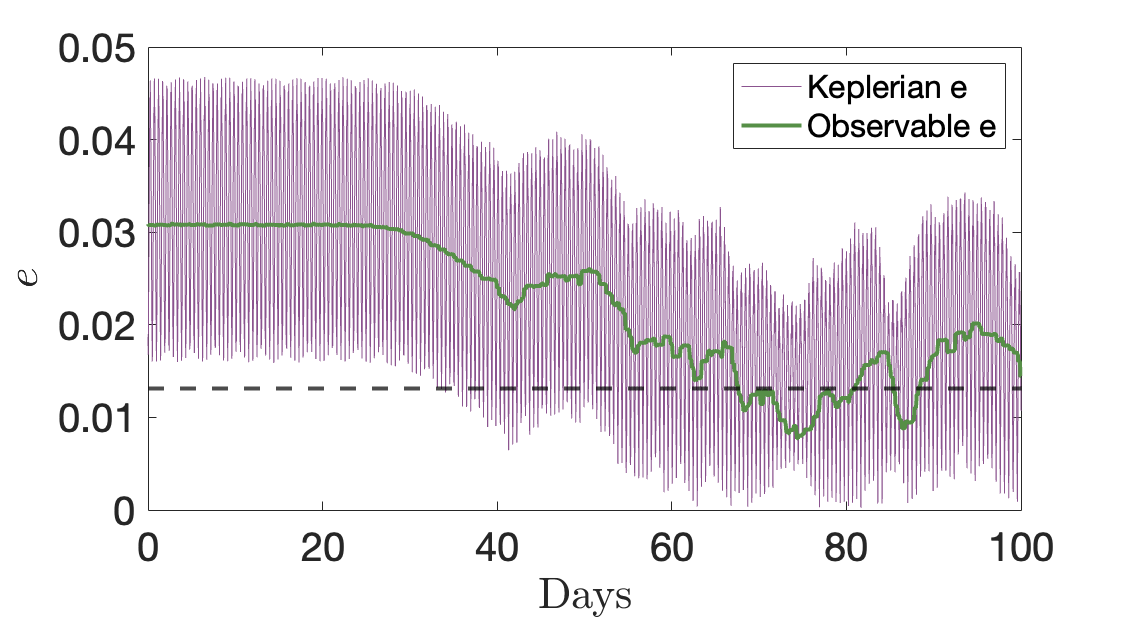} 
   \caption{The Keplerian and observable eccentricity of a system that begins tumbling. As the secondary enters a tumbling state, the eccentricity decreases as the orbital angular momentum decreases. The dashed line corresponds to the equilibrium Keplerian value.}
   \label{tumble_e}
\end{figure}

\begin{figure}[ht!]
   \centering
   \includegraphics[width = 3in]{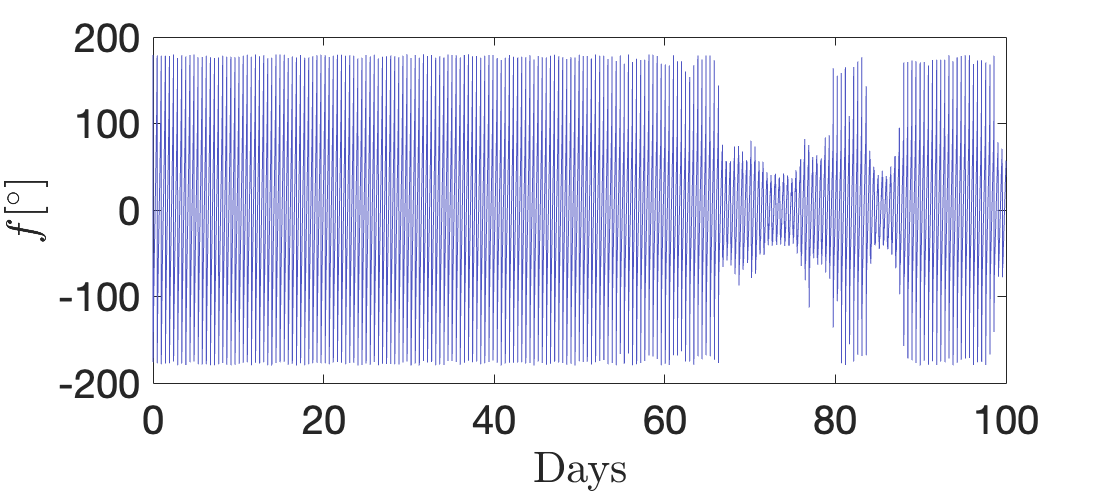} 
   \caption{The true anomaly of a system that begins tumbling. At some threshold, the true anomaly switches from clocking to librating, then back to clocking as angular momentum is exchanged between the orbit and secondary. Comparing to Fig.\ \ref{tumble_h}, this threshold is near when the orbital angular momentum crosses its nearest equilibrium value, or equivalently when the osculating eccentricity decreases to its equilibrium value, as seen in Fig.\ \ref{tumble_e}.}
   \label{tumble_f}
\end{figure}

As the secondary begins tumbling, the specific orbit angular momentum increases to its equilibrium value. At the same time, the Keplerian eccentricity decreases to its equilibrium value, as shown in Fig.\ \ref{tumble_e}, where the dashed black line is the calculated equilibrium eccentricity from Eq.\ \ref{kep_ecc}. The time period when the specific orbital angular momentum is equal to its equilibrium value is roughly the same time period when the Keplerian eccentricity is equal to its equilibrium value. This is also roughly the same time period when the true anomaly, shown in Fig.\ \ref{tumble_f}, returns to a state of libration, before once again tracking when the orbital angular momentum again decreases and the osculating eccentricity increases.

Thus, we have demonstrated how the onset of tumbling in the secondary can have significant effects on the orbit dynamics, a general result beyond the application to Didymos and DART. We note this decrease in eccentricity is not caused by enhanced dissipation (e.g. \cite{quillen2020excitation}), which is not modeled here, but by the commensurability between the secondary and orbit angular momenta. The angular momentum lost by the secondary is sufficient to directly decrease the orbit's eccentricity. As the secondary begins tumbling, the decrease in eccentricity is substantial and rapid as the orbit becomes more circular and settles into the nearest equilibrium orbit. Thus, the onset of tumbling in the secondary may be observable from the ground in lightcurve data. As the system continues to evolve, this exchange in angular momentum can continue, causing the eccentricity to fluctuate. Furthermore, the true anomaly can then switch between epochs of circulating and librating, depending on the orbit's angular momentum.

\subsection{Orbit Period}
As discussed in \cite{meyer2021libration}, variations in the stroboscopic orbit period can also provide clues to the secondary's spin state. While the secondary is in stable libration, the orbit period fluctuates with the same period as the orbit's apsidal precession. There are also smaller fluctuations, which appear to be driven by the short-period apsidal precession and libration periods, and may be influenced by other frequencies as well. For the preliminary Dimorphos shape, which is nearly axisymmetric \citep{daly2023}, there isn't a clear dominant frequency in the short-period fluctuations, unlike the more elongated shapes investigated in \cite{meyer2021libration}. Unfortunately, these smaller fluctuations are likely not observable in lightcurve date \citep{meyer2021libration}, but may be observable by Hera.

\begin{figure}[ht!]
   \centering
   \includegraphics[width = 3in]{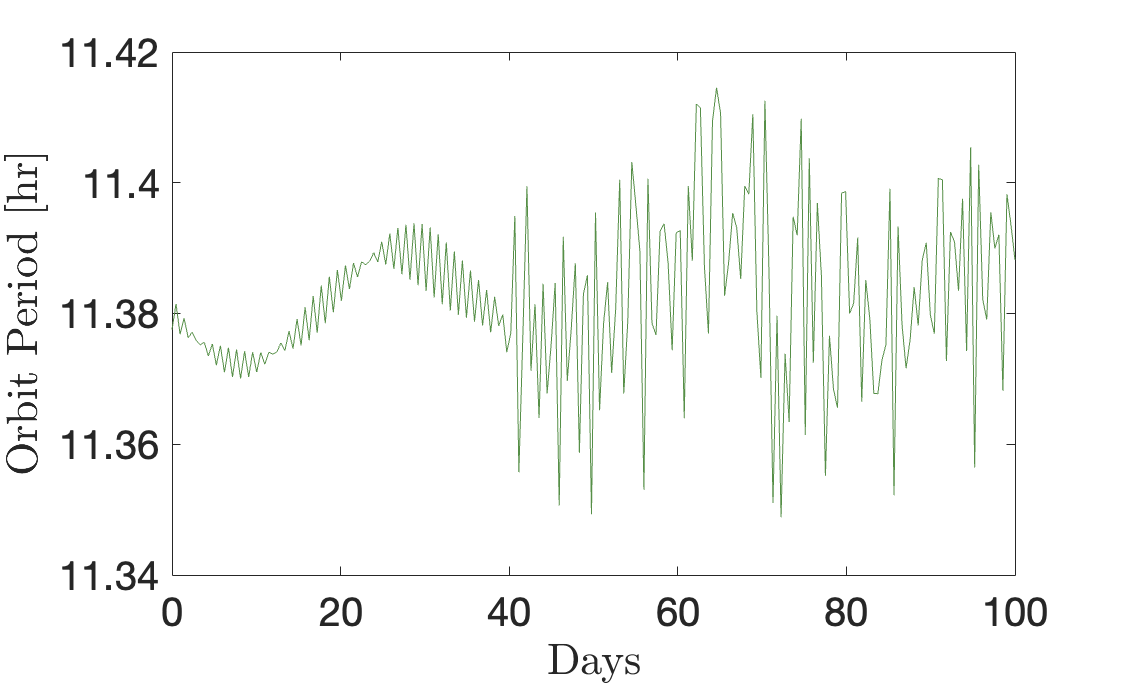} 
   \caption{Stroboscopic orbit period over time for the tumbling case study in the previous section. Prior to the onset of tumbling, the orbit period experiences small, periodic variations. As the secondary begins to tumble, the amplitude of orbit period variations increases and the periodicity is absent.}
   \label{tumble_period}
\end{figure}

The stroboscopic orbit period for the tumbling case study is plotted in Fig.\ \ref{tumble_period} over time. We see the expected orbit period variations dominated by the orbit's precession prior to tumbling, but as the secondary enters NPA rotation, the amplitude of orbit period variations increases significantly and the periodicity is lost. This implies that, along with a decrease in observed eccentricity, large deviations from expected mutual event timings in lightcurves can also indicate tumbling. Thus, we have established two independent indications of tumbling that are observable in lightcurve observations: a rapid decrease in the observable eccentricity followed by fluctuations and large, aperiodic variations in the stroboscopic orbit period. So even if the spin state of the secondary is not directly observable from the ground, it is still possible to predict if it is tumbling or not from other features in the lightcurves. However, given observational errors and noise, it may be difficult to detect these characteristics in ground-based lightcurves.

\section{Dimorphos Mass Loss and Reshaping}
\label{sec:reshaping} 
We now relax our assumption of no mass loss or reshaping of Dimorphos from the DART impact. This is an important consideration, as \cite{nakano2022nasa} showed that reshaping of Dimorphos can contribute to the momentum enhancement factor estimate. To achieve this, we modify our algorithm so that after the system's mass is calculated to match the pre-impact orbit period, we re-sample the secondary size and shape, keeping its density the same. DART impacted roughly along the intermediate axis of Dimorphos \citep{daly2023}, so any reshaping will decrease the ellipsoid's b axis to first order. This is equivalent to increasing the a/b ratio and decreasing the b/c ratio. In reality, reshaping may result in an asymmetric secondary \citep{raducan2022global}. However, while our secondary is still an ellipsoid shape, we are in effect changing its moments of inertia, which are the dominant quantities in calculating the mutual potential.

We re-sample a/b and b/c for Dimorphos, raising the upper limit of a/b to 1.3 and decreasing the lower limit of b/c to 1.1. To reflect the reality of decreasing the b axis, in re-sampling the axis ratios, we adjust the uniform distribution so that the lower limit of a/b is equal to its pre-impact value (so it always increases) and the upper limit of b/c is equal to its pre-impact value (so it always decreases). The new limits on post-impact a/b and b/c are somewhat arbitrary but allow for relatively large changes in the secondary shape.

To account for mass loss, we also re-sample the volume-equivalent radius of Dimorphos. We center our sampling on the pre-impact value and draw from a half-normal distribution with a standard deviation of 0.1 m, ensuring we do not increase the radius. The standard deviation of the post-impact radius was chosen to approximately match the initial mass-loss estimate of \cite{graykowski2023}, around 0.3-0.5\% of the assumed Dimorphos pre-impact mass.

\begin{figure}[ht!]
   \centering
   \includegraphics[width = 3in]{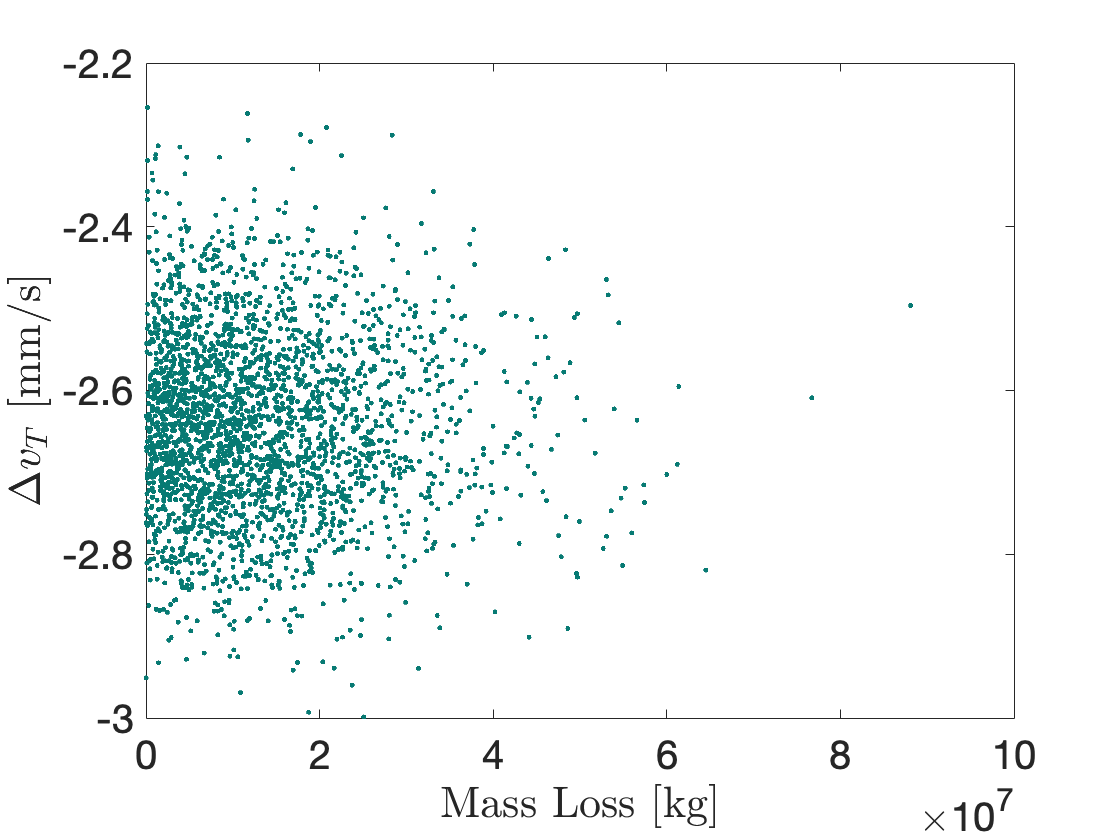} 
   \caption{Calculated $\Delta v_T$ as a function of Dimorphos's mass loss. Within reasonable values for mass loss, there is no strong trend in $\Delta v_T$.}
   \label{delta_v_massloss}
\end{figure}

\begin{figure}[ht!]
   \centering
   \includegraphics[width = 3in]{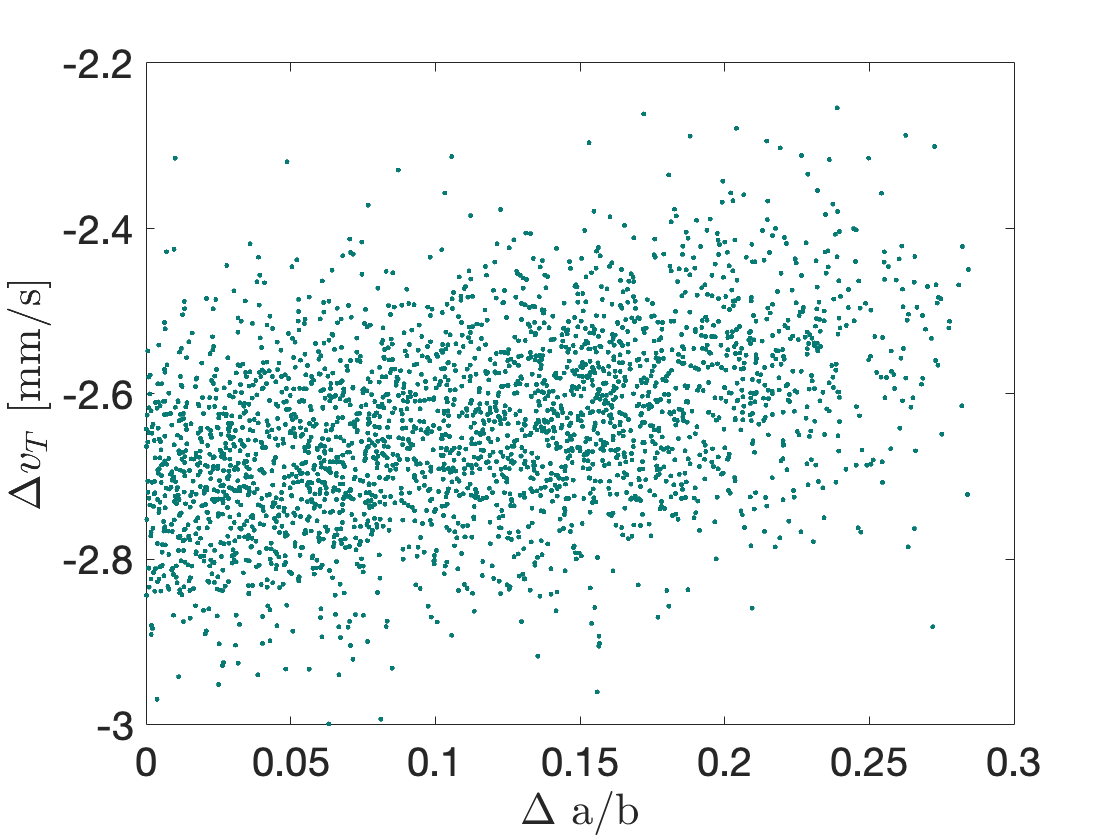} 
   \caption{Calculated $\Delta v_T$ as a function of Dimorphos's reshaping of its a/b ratio. For impacts that cause more reshaping, the required $\Delta v_T$ magnitude to achieve the post-impact orbit period is smaller on average.}
   \label{delta_v_reshape}
\end{figure}

Accounting for mass loss and reshaping slightly reduces the magnitude of $\Delta v_T$ needed to achieve the post-impact orbit period. Our new estimate is $\Delta v_T \approx -2.6\pm0.1$ mm/s. This effect was not considered in \cite{cheng2023}, and thus they may have over-approximated the impact-induced $\Delta v_T$ if secondary reshaping is significant. To explore this shift, we plot $\Delta v_T$ as a function of the mass loss in Fig.\ \ref{delta_v_massloss} and as a function of reshaping a/b in Fig.\ \ref{delta_v_reshape}. We find $\Delta v_T$ has a negligible dependence on the mass loss, likely given the minimal percent change of the mass. However, the magnitude of $\Delta v_T$ slightly decreases with larger changes to a/b, consistent with results from \cite{nakano2022nasa}, who showed reshaping of the secondary changes the mutual potential between the bodies, which contributes to changes in the orbit period. We find no trend with $\Delta v_T$ and the b/c ratio.

\begin{figure}[ht!]
   \centering
   \includegraphics[width = 3in]{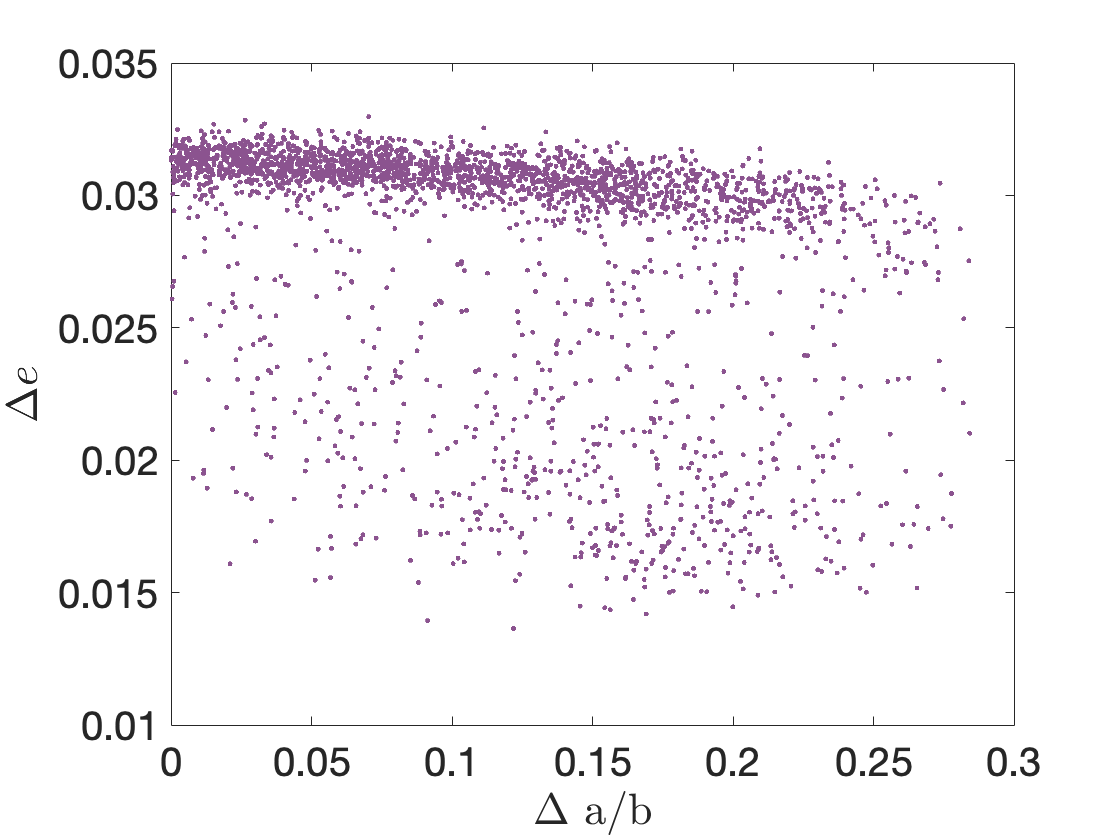} 
   \caption{Calculated observable $\Delta e$ as a function of Dimorphos's reshaping of its a/b ratio. For impacts that cause more reshaping, the change in eccentricity is on-average reduced.}
   \label{delta_e_reshape}
\end{figure}

We also find a relationship between reshaping of a/b and the post-impact observable eccentricity. For larger changes in a/b, the change in observable eccentricity slightly decreases on average. This is another important consideration to make when interpreting the system's current eccentricity. Fig.\ \ref{delta_e_reshape} shows this relationship.

\begin{figure}[ht!]
   \centering
   \includegraphics[width = 3in]{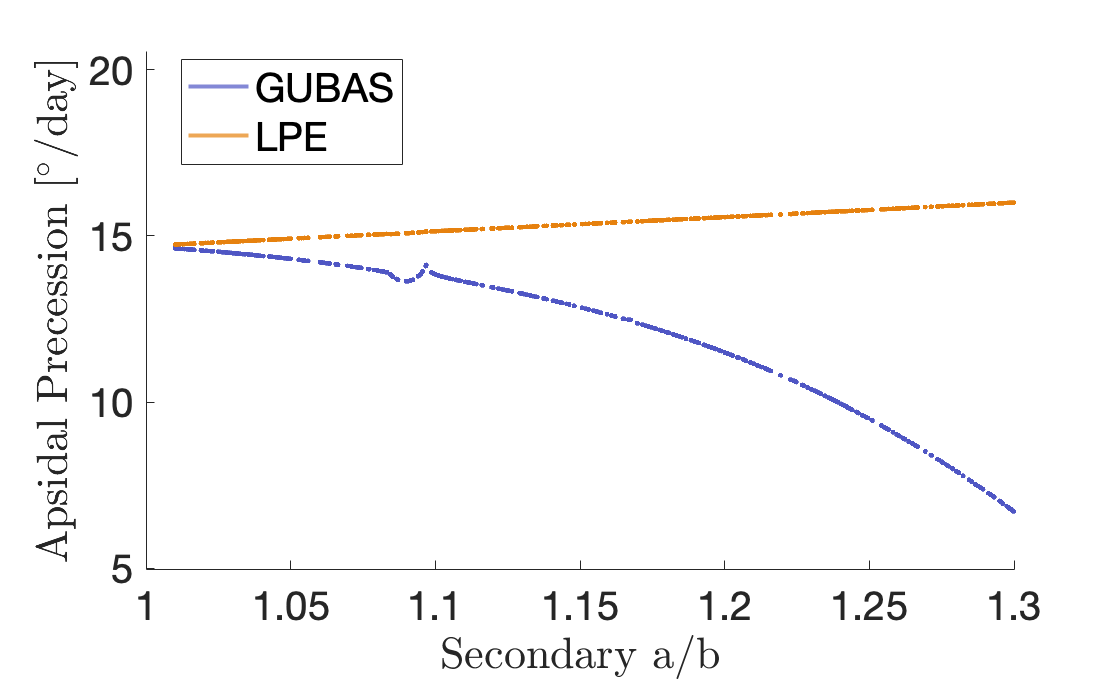} 
   \caption{The precession rate as a function of the secondary's elongation a/b, calculated using the F2BP dynamics (GUBAS) and the analytical Lagrange Planetary Equations (LPEs) with the nominal primary $J_2$. Including secondary libration in the dynamics results in a decrease in the apsidal precession rate, whereas ignoring it results in an increase. The effect of a resonance is visible near a/b$=1.1$, discussed in detail in \cite{agrusa2021excited}.}
   \label{precession_libration}
\end{figure}

While accounting for mass loss and reshaping slightly reduces the $\Delta v_T$ and $\Delta e$ estimate, there is no statistical difference in the post-impact observable semimajor axis. This is expected, as this quantity is driven by the estimated pre-impact observable semimajor axis and the orbit period change, which have been directly estimated. Reshaping also increases the chances of the secondary entering a state of NPA rotation, with roughly 40\% of our simulations experiencing tumbling when reshaping is considered, compared to about 18\% when reshaping is ignored. Allowing for reshaping also reduces the lower limit of the expected post-impact precession rate, so that the precession rate now lies in the range $4-20^\circ$/day. Thus, elongation of the secondary reduces the precession rate of the orbit, as described in \cite{cuk2010orbital}. This is especially interesting, since analytical expressions (e.g. Eq. \ref{LPEprecession}) predict a more elongated secondary increases the orbit's precession rate. Thus, the secondary's libration is a key component to accurately calculating the precession rate of a binary asteroid. This is illustrated in Fig. \ref{precession_libration}, where the precession rate of the nominal system is calculated using the F2BP dynamics in \textsc{gubas} and compared to the analytic calculation using the LPEs. A detailed analysis of the role of secondary shape and libration on the system's precession rate is left as a future investigation.

As pointed out by \cite{nakano2022nasa}, the adjustment to $\Delta v_T$ caused by reshaping could affect the estimate of the momentum enhancement factor calculated by \cite{cheng2023}. However, given the current uncertainties in the ejecta momentum direction, this small adjustment to $\Delta v_T$ will not have a noticeable affect on their results.

\section{Discussion and Conclusion}
\label{sec:conclusion} 
Due to spin-orbit coupling, the dynamics of close, irregularly shaped near-Earth binary asteroids, such as Didymos, are non-Keplerian. This makes describing the orbit difficult when such differences are of consequence, as traditional Keplerian elements are misleading, particularly when the true anomaly is in a state of libration. To handle this difficulty, we introduce a set of observable elements to describe the semimajor axis and eccentricity that an external observer would see. This allows us to match dynamical simulations to real-world observations. The stroboscopic orbit period is also useful for this, as it parallels mutual event timings used in lightcurve measurements to calculate the mutual orbit period.

In this work, we developed novel analytical expressions to determine when a perturbation to a binary asteroid's orbit is sufficient to break the equilibrium configuration and allow the true anomaly to circulate rather than librate. This threshold occurs when the semimajor axis is either increased or decreased by the difference between the Keplerian and observable semimajor axes of an equilibrated system. This is directly applicable to the DART impact, for which our calculation exceeded this criterion by roughly $50\%$, leading to circulation of the true anomaly in the post-impact Didymos orbit. This means analytical averaged Lagrange Planetary Equations can be used to reliably estimate the system's precession rate, provided the secondary has only minimal elongation.

In studying the post-impact Didymos mutual orbit, we used a similar algorithm as that used by \cite{cheng2023} to calculate the system mass and the tangential $\Delta v_T$ imparted by the impact. We show that using Keplerian dynamics overestimates the mass by around 1--3\%. However, the uncertainty in the pre-impact semimajor axis is much larger than this error. Indeed, the current system is dominated by uncertainties in the pre-impact semimajor axis. 

Owing to the large uncertainties in the system, particularly on the semimajor axis, our estimate of the system's bulk density of $2.4\pm0.3$ g/cm$^3$ ($1\sigma$) matches the estimate in \cite{daly2023} despite the errors in a Keplerian approach. This is because the mass uncertainty is dominated by the unknown semimajor axis. However, as a function of the semimajor axis, our mass estimate is systematically different from the Keplerian approach, as shown in Fig. \ref{f2bp_mass}.  In calculating $\Delta v_T$ for the impact, we obtain an estimate of $-2.7\pm0.1$ mm/s ($1\sigma$), matching the results from \cite{cheng2023}. However, for both these quantities, we find a strong dependence on the pre-impact observable semimajor axis. Using numerical fits based on Keplerian relationships we find $\rho \approx 2.44\left(\frac{a_0}{1206\text{ m}}\right)^3-0.04\text{ g/cm}^3$ and $\Delta v_T \approx -5.86\sqrt{\frac{a_0}{1206\text{ m}}}+3.19\text{ mm/s}$, where $a_0$ is the pre-impact observable semimajor axis in meters. We also find any component of $\Delta v$ misaligned with the orbit tangent direction does not have a noticeable effect on the post-impact orbit period.

In this work we also calculate the change in observable semimajor axis and eccentricity caused by the impact, which was not included in previous analyses of the DART impact. We find $\Delta a=-37\pm1$ m ($1\sigma$), with a linear dependence on the pre-impact observable semimajor axis of $\Delta a \approx -37.64\left(\frac{a_0}{1206\text{ m}}\right)+0.91\text{ m}$ for $a_0$ in meters. This provides an estimate of the post-impact observable semimajor axis of $1170\pm34$ m ($1\sigma$). For the observable eccentricity, we estimate $0.031\pm0.002$. However, the presence of pre-impact eccentricity can appreciably increase or decrease the post-impact eccentricity, depending on where in the secondary's orbit the perturbation occurs. Reshaping of Dimorphos by the DART impact may also reduce the observable eccentricity.

As we point out in this work, the strong dependence of the density, velocity change, and post-impact semimajor axis on the pre-impact observable semimajor axis suggests this is a quantity of particular interest for the ESA Hera mission to Didymos \citep{michel2022esa}. By measuring the post-impact observable semimajor axis, the pre-impact observable semimajor axis can be inferred, leading to a much narrower estimate of $\Delta v_T$. However, this may become more complicated if secular effects have noticeably evolved the semimajor axis in the time between the impact and Hera's arrival \citep{meyer2023energy}. No close encounters that could perturb the Didymos mutual orbit further are expected with any of the terrestrial planets for thousands of years, using the semi-analytical propagation of the heliocentric orbit from \cite{Fuentes-Munoz_2022}.

The observable eccentricity offers an interesting method of predicting the secondary's attitude stability. As demonstrated in this work, the onset of tumbling in the secondary can rapidly decrease the orbit's eccentricity. Thus, an evolving eccentricity is indicative of a tumbling secondary, whereas a constant eccentricity suggests stable libration. Consistent with \cite{meyer2021libration}, it is possible to see evidence of tumbling in variations in the stroboscopic orbit period. The secondary entering a state of tumbling can cause the orbit to return to an equilibrium state where the true anomaly librates and the longitude of periapsis tracks the secondary. However, as the secondary's spin state evolves, the orbit can leave this state again, allowing the true anomaly to return to circulation. This illustrates the angular momentum exchange between the secondary and the orbit in a perturbed binary asteroid. However, given observational errors and noise, the absence of these signals in lightcurves does not rule out the possibility of tumbling. Nevertheless, this analysis can provide context for future observations to aid in interpreting the data and has general applicability beyond just the Didymos system.

Lastly, we discuss the implications of allowing for mass loss and reshaping of Dimorphos caused by the DART impact. The observable semimajor axis is unaffected by either mass loss or reshaping, but the presence of reshaping can reduce the estimates of $\Delta v_T$ and $\Delta e$. Specifically, increasing a/b of Dimorphos leads to a smaller magnitude of velocity change required to achieve the measured post-impact orbit period, consistent with findings from \cite{nakano2022nasa}. Increasing a/b also decreases the post-impact observable eccentricity and precession rate, the latter of which points to a trade-off between primary oblateness and secondary elongation in the mutual orbit precession. Thus, the current post-impact shape of Dimorphos that the Hera mission will measure will allow for the most accurate calculation of $\Delta v_T$ and $\Delta e$.

\section*{Acknowledgements}
This work was supported by the DART mission, NASA Contract 80MSFC20D0004. P.M. acknowledges funding support from CNES, from the European Union’s Horizon 2020 research and innovation programme under grant agreement No 870377 (project NEO-MAPP), from ESA and from the University of Tokyo. The work of S.R.C. and S.P.N.  was conducted at the Jet Propulsion Laboratory under a contract with the National Aeronautics and Space Administration. I.G. and K.T. acknowledge support from the European Union’s Horizon 2020 research and innovation programme under grant agreement No 870377 (project NEO-MAPP). The work by P.P. and P.S. was supported by the Grant Agency of the Czech Republic, grant 20-04431S.

 \section*{Notation Appendix}

 \begin{table}[ht]
 \caption{\label{variables} Notation and definitions for the variables used in this paper.}
 \begin{center}
  \begin{tabular}{ |c|c| } 
  \hline
  Notation & Definition  \\
  \hline
  $\Delta \vec{v}$ & Change in velocity vector of the secondary \\
  $\Delta v_T$ & Change in tangential velocity of the secondary\\
  a, b, c & Semiaxes of triaxial ellipsoid shapes \\
  $\mu$ & Standard gravitational parameter of binary system \\
  $r$ & Separation distance of primary and secondary \\
  $\bar{I}_A$ & Mass-normalized principal inertia of primary \\
  $\bar{I}_B$ & Mass-normalized principal inertia of secondary \\
  $h^*$ & Equilibrium specific orbital angular momentum \\
  $h_{B}$ & Specific angular momentum of secondary \\
  $h_{orb}$ & Specific orbit angular momentum \\
  $\Delta h$ & Perturbation to specific orbit angular momentum \\
  $\dot{\theta}$ & Orbit rate of secondary \\
  $\dot{\theta}^*$ & Equilibrium orbit rate of secondary \\
  $e_{kep}$ & Keplerian eccentricity \\
  $a_{kep}$ & Keplerian semimajor axis \\
  $e^*_{kep}$ & Equilibrium Keplerian eccentricity \\
  $a^*_{kep}$ & Equilibrium Keplerian semimajor axis \\
  $a$ & Observable semimajor axis \\
  $e$ & Observable eccenticity \\
  $f$ & True anomaly of secondary \\
  $M$ & Mean anomaly of secondary \\
  $\bar{\omega}$ & Longitude of periapsis of secondary \\
  $\omega$ & Argument of periapsis of secondary \\
  $\Omega$ & Longitude of the ascending node of secondary \\
  $\rho$ & Bulk density \\
  $\Delta T$ & Error between measured and simulated orbit period\\
  $P_1$ & Pre-impact measured orbit period of secondary \\
  $a_1$ & Pre-impact observable semimajor axis of secondary\\
  $P_2$ & Post-impact measured orbit period of secondary \\
  $a_2$ & Post-impact observable semimajor axis of secondary\\
  $\theta_1$,$\theta_2$,$\theta_3$ & 1-2-3 (roll-pitch-yaw) Euler angles of the secondary\\
   \hline
  \end{tabular}
  \end{center}
  \end{table}


\bibliography{_bib}{}

\begin{thebibliography}{}
\expandafter\ifx\csname natexlab\endcsname\relax\def\natexlab#1{#1}\fi
\providecommand{\url}[1]{\href{#1}{#1}}
\providecommand{\dodoi}[1]{doi:~\href{http://doi.org/#1}{\nolinkurl{#1}}}
\providecommand{\doeprint}[1]{\href{http://ascl.net/#1}{\nolinkurl{http://ascl.net/#1}}}
\providecommand{\doarXiv}[1]{\href{https://arxiv.org/abs/#1}{\nolinkurl{https://arxiv.org/abs/#1}}}

\bibitem[{Agrusa {et~al.}(2020)Agrusa, Richardson, Davis, Fahnestock,
  Hirabayashi, Chabot, Cheng, Rivkin, Michel, Group,
  {et~al.}}]{agrusa2020benchmarking}
Agrusa, H.~F., Richardson, D.~C., Davis, A.~B., {et~al.} 2020, Icarus, 349,
  113849

\bibitem[{Agrusa {et~al.}(2021)Agrusa, Gkolias, Tsiganis, Richardson, Meyer,
  Scheeres, {\'C}uk, Jacobson, Michel, Karatekin, {et~al.}}]{agrusa2021excited}
Agrusa, H.~F., Gkolias, I., Tsiganis, K., {et~al.} 2021, Icarus, 370, 114624

\bibitem[{Bellerose \& Scheeres(2008)}]{bellerose2008energy}
Bellerose, J., \& Scheeres, D.~J. 2008, Celestial Mechanics and Dynamical
  Astronomy, 100, 63

\bibitem[{Borderies \& Yoder(1990)}]{borderies1990phobos}
Borderies, N., \& Yoder, C. 1990, Astronomy and Astrophysics (ISSN 0004-6361),
  vol. 233, no. 1, July 1990, p. 235-251., 233, 235

\bibitem[{Borderies-Rappaport \& Longaretti(1994)}]{borderies1994test}
Borderies-Rappaport, N., \& Longaretti, P.-Y. 1994, Icarus, 107, 129

\bibitem[{Cheng {et~al.}(2023)Cheng, Agrusa, Barbee, Meyer, Farnham, Raducan,
  Richardson, Dotto, Zinzi, Della~Corte, {et~al.}}]{cheng2023}
Cheng, A.~F., Agrusa, H.~F., Barbee, B.~W., {et~al.} 2023, Nature, 1

\bibitem[{{\'C}uk {et~al.}(2021){\'C}uk, Jacobson, \& Walsh}]{cuk2021barrel}
{\'C}uk, M., Jacobson, S.~A., \& Walsh, K.~J. 2021, The Planetary Science
  Journal, 2, 231

\bibitem[{{\'C}uk \& Nesvorn{\`y}(2010)}]{cuk2010orbital}
{\'C}uk, M., \& Nesvorn{\`y}, D. 2010, Icarus, 207, 732

\bibitem[{Daly {et~al.}(2023)Daly, Ernst, Barnouin, Chabot, Rivkin, Cheng,
  Adams, Agrusa, Abel, Alford, {et~al.}}]{daly2023}
Daly, R.~T., Ernst, C.~M., Barnouin, O.~S., {et~al.} 2023, Nature, 1

\bibitem[{Davis \& Scheeres(2020)}]{davis2020doubly}
Davis, A.~B., \& Scheeres, D.~J. 2020, Icarus, 341, 113439

\bibitem[{Duboshin(1958)}]{duboshin1958differential}
Duboshin, G. 1958, Soviet Astronomy, Vol. 2, p. 239, 2, 239

\bibitem[{Fahnestock \& Scheeres(2008)}]{fahnestock2008simulation}
Fahnestock, E.~G., \& Scheeres, D.~J. 2008, Icarus, 194, 410

\bibitem[{Fuentes-Muñoz {et~al.}(2022)Fuentes-Muñoz, Meyer, \&
  Scheeres}]{Fuentes-Munoz_2022}
Fuentes-Muñoz, O., Meyer, A.~J., \& Scheeres, D.~J. 2022, The Planetary
  Science Journal, 3, 257, \dodoi{10.3847/PSJ/ac83c6}

\bibitem[{Graykowski {et~al.}(2023)Graykowski, Lambert, Marchis, Cazeneuve,
  Dalba, Esposito, Peluso, Sgro, Blaclard, Borot, {et~al.}}]{graykowski2023}
Graykowski, A., Lambert, R.~A., Marchis, F., {et~al.} 2023, Nature, 1

\bibitem[{Ho {et~al.}(2023)Ho, Wold, Poursina, \& Conway}]{ho2023accuracy}
Ho, A., Wold, M., Poursina, M., \& Conway, J.~T. 2023, arXiv preprint
  arXiv:2301.10083

\bibitem[{Hou {et~al.}(2017)Hou, Scheeres, \& Xin}]{hou2017mutual}
Hou, X., Scheeres, D.~J., \& Xin, X. 2017, Celestial Mechanics and Dynamical
  Astronomy, 127, 369

\bibitem[{Jacobson \& Scheeres(2011)}]{jacobson2011long}
Jacobson, S.~A., \& Scheeres, D.~J. 2011, The Astrophysical Journal Letters,
  736, L19

\bibitem[{Jafari-Nadoushan(2023)}]{jafari2023surfing}
Jafari-Nadoushan, M. 2023, Monthly Notices of the Royal Astronomical Society,
  520, 3514

\bibitem[{Maciejewski(1995)}]{maciejewski1995reduction}
Maciejewski, A.~J. 1995, Celestial Mechanics and Dynamical Astronomy, 63, 1

\bibitem[{McMahon \& Scheeres(2013)}]{mcmahon2013dynamic}
McMahon, J.~W., \& Scheeres, D.~J. 2013, Celestial Mechanics and Dynamical
  Astronomy, 115, 365

\bibitem[{Meyer {et~al.}(2021{\natexlab{a}})Meyer, Scheeres, Naidu, Benner,
  Pravec, \& Scheirich}]{meyer2021modeling}
Meyer, A., Scheeres, D., Naidu, S., {et~al.} 2021{\natexlab{a}}, AAS/Division
  of Dynamical Astronomy Meeting, 53, 405

\bibitem[{Meyer {et~al.}(2023)Meyer, Scheeres, Agrusa, Noiset, McMahon,
  Karatekin, Hirabayashi, \& Nakano}]{meyer2023energy}
Meyer, A.~J., Scheeres, D.~J., Agrusa, H.~F., {et~al.} 2023, Icarus, 391,
  115323

\bibitem[{Meyer {et~al.}(2021{\natexlab{b}})Meyer, Gkolias, Gaitanas, Agrusa,
  Scheeres, Tsiganis, Pravec, Benner, Ferrari, \& Michel}]{meyer2021libration}
Meyer, A.~J., Gkolias, I., Gaitanas, M., {et~al.} 2021{\natexlab{b}}, The
  planetary science journal, 2, 242

\bibitem[{Michel {et~al.}(2022)Michel, K{\"u}ppers, Bagatin, Carry, Charnoz,
  De~Leon, Fitzsimmons, Gordo, Green, H{\'e}rique, {et~al.}}]{michel2022esa}
Michel, P., K{\"u}ppers, M., Bagatin, A.~C., {et~al.} 2022, The Planetary
  Science Journal, 3, 160

\bibitem[{Moeckel(2018)}]{moeckel2018counting}
Moeckel, R. 2018, Celestial Mechanics and Dynamical Astronomy, 130, 17

\bibitem[{Naidu {et~al.}(2020)Naidu, Benner, Brozovic, Nolan, Ostro, Margot,
  Giorgini, Hirabayashi, Scheeres, Pravec, {et~al.}}]{naidu2020radar}
Naidu, S., Benner, L., Brozovic, M., {et~al.} 2020, Icarus, 348, 113777

\bibitem[{Naidu {et~al.}(2022)Naidu, Chesley, Farnocchia, Moskovitz, Pravec,
  Scheirich, Thomas, \& Rivkin}]{naidu2022anticipating}
Naidu, S.~P., Chesley, S.~R., Farnocchia, D., {et~al.} 2022, The Planetary
  Science Journal, 3, 234

\bibitem[{Naidu \& Margot(2015)}]{naidu2015near}
Naidu, S.~P., \& Margot, J.-L. 2015, The Astronomical Journal, 149, 80

\bibitem[{Nakano {et~al.}(2022)Nakano, Hirabayashi, Agrusa, Ferrari, Meyer,
  Michel, Raducan, S{\'a}nchez, \& Zhang}]{nakano2022nasa}
Nakano, R., Hirabayashi, M., Agrusa, H.~F., {et~al.} 2022, The Planetary
  Science Journal, 3, 148

\bibitem[{Pravec {et~al.}(2016)Pravec, Scheirich, Ku{\v{s}}nir{\'a}k, Hornoch,
  Gal{\'a}d, Naidu, Pray, Vil{\'a}gi, Gajdo{\v{s}}, Korno{\v{s}},
  {et~al.}}]{pravec2016binary}
Pravec, P., Scheirich, P., Ku{\v{s}}nir{\'a}k, P., {et~al.} 2016, Icarus, 267,
  267

\bibitem[{Quillen {et~al.}(2022)Quillen, LaBarca, \& Chen}]{quillen2022non}
Quillen, A.~C., LaBarca, A., \& Chen, Y. 2022, Icarus, 374, 114826

\bibitem[{Quillen {et~al.}(2020)Quillen, Lane, Nakajima, \&
  Wright}]{quillen2020excitation}
Quillen, A.~C., Lane, M., Nakajima, M., \& Wright, E. 2020, Icarus, 340, 113641

\bibitem[{Raducan \& Jutzi(2022)}]{raducan2022global}
Raducan, S.~D., \& Jutzi, M. 2022, The planetary science journal, 3, 128

\bibitem[{Renner \& Sicardy(2006)}]{renner2006use}
Renner, S., \& Sicardy, B. 2006, Celestial Mechanics and Dynamical Astronomy,
  94, 237

\bibitem[{Richardson {et~al.}(2022)Richardson, Agrusa, Barbee, Bottke, Cheng,
  Eggl, Ferrari, Hirabayashi, Karatekin, McMahon,
  {et~al.}}]{richardson2022predictions}
Richardson, D.~C., Agrusa, H.~F., Barbee, B., {et~al.} 2022, The Planetary
  Science Journal, 3, 157

\bibitem[{Scheeres(2006)}]{scheeres2006relative}
Scheeres, D.~J. 2006, Celestial Mechanics and Dynamical Astronomy, 94, 317

\bibitem[{Scheeres(2009)}]{scheeres2009stability}
---. 2009, Celestial Mechanics and Dynamical Astronomy, 104, 103

\bibitem[{Tan {et~al.}(2023)Tan, Wang, \& Hou}]{tan2023attitude}
Tan, P., Wang, H.-s., \& Hou, X.-y. 2023, Icarus, 390, 115289

\bibitem[{Thomas {et~al.}(2023)Thomas, Naidu, Scheirich, Moskovitz, Pravec,
  Chesley, Rivkin, Osip, Lister, Benner, {et~al.}}]{thomas2023}
Thomas, C.~A., Naidu, S.~P., Scheirich, P., {et~al.} 2023, Nature, 1

\bibitem[{Wang \& Hou(2021)}]{wang2021break}
Wang, H.-S., \& Hou, X.-Y. 2021, Monthly Notices of the Royal Astronomical
  Society, 505, 6037

\end{thebibliography}
\bibliographystyle{aasjournal}



\end{document}